\documentclass[letterpaper, 10 pt, conference]{ieeeconf}

\usepackage{amsmath} % assumes amsmath package installed
\usepackage{amssymb}  % assumes amsmath package installed
\usepackage{subfig}
\usepackage[ruled,longend]{algorithm2e}
\usepackage{multicol}
\usepackage{color}
\usepackage{xcolor}
\usepackage{hyperref}
\hypersetup{colorlinks,linkcolor=blue}
\usepackage{threeparttable}
\usepackage[export]{adjustbox}
\usepackage{graphicx} 
\usepackage{caption}%,subcaption}

\newcommand{\T}{\ensuremath{^\mathsf{T}}}
\usepackage{scalerel}

\newcommand\reallywidehat[1]{\arraycolsep=0pt\relax%
\begin{array}{c}
\stretchto{
  \scaleto{
    \scalerel*[\widthof{\ensuremath{#1}}]{\kern-.5pt\bigwedge\kern-.5pt}
    {\rule[-\textheight/2]{1ex}{\textheight}} %WIDTH-LIMITED BIG WEDGE
  }{\textheight} % 
}{0.5ex}\\           % THIS SQUEEZES THE WEDGE TO 0.5ex HEIGHT
#1\\                 % THIS STACKS THE WEDGE ATOP THE ARGUMENT
\rule{-1ex}{0ex}
\end{array}
}

\begin{document}

\title{Optimal Control of Material Micro-Structures}

\author{Aayushman Sharma, Zirui Mao, Haiying Yang, Suman Chakravorty, Michael J Demkowicz, Dileep Kalathil}

\maketitle

\begin{abstract}
In this paper, we consider the optimal control of material micro-structures. Such material micro-structures are modeled by the so-called phase field model. We study the underlying physical structure of the model and propose a data based approach for its optimal control, along with a comparison to the control using a state of the art Reinforcement Learning (RL) algorithm. Simulation results show the feasibility of optimally controlling such micro-structures to attain desired material properties and complex target micro-structures.

\end{abstract}

\section{Introduction}

% The optimal control of the non-linear dynamical system with continuous state and action space is a difficult problem due to the inherent `curse of dimensionality' on dynamic programming problems.
In this paper, we consider the optimal control of complex and high DoF (Degree of Freedom) material micro-structure dynamics, by doing which we aim at developing a robust tool for exploring the fundamental limits of materials micro-structure control during processing. The micro-structure dynamics process is governed by the so-called  Phase Field (PF) model  \cite{Nikolas2010}, which is a powerful tool for modeling micro-structure evolution of materials. It represents the spatial distribution of physical quantities of interest by an order parameter field governed by one or a set of partial differential equations (PDEs). The Phase Field method can naturally handle material systems with complicated interface, thanks to which it has been successfully applied to the simulation of a wide range of materials micro-structure evolution processes, such as \textit{e.g.,} alloy solidification \cite{Boe2002}\cite{MOELANS2008268}, alloy phase transformation \cite{NGUYEN2016322}\cite{Rui2011}, grain growth\cite{Flint2019}\cite{CHENIOUR2020152069}, and failure mechanism of materials \cite{KHADERI20141}\cite{HANSENDORR201925} \textit{etc}.\\
%\subsection{Related Work}

The phase field model is complex, nonlinear, and high DOF, and does not admit analytical solutions. Thus, data based approaches are natural to consider for the control of these systems.
In recent work \cite{D2C1.0}, we proposed a novel decoupled data based control (D2C) algorithm for learning to control an unknown nonlinear dynamical system. Our approach introduced a rigorous decoupling of the open-loop (planning) problem from the closed-loop (feedback control) problem. This decoupling allows us to come up with a highly efficient approach to solve the problem in a completely data based fashion. Our approach proceeds in two steps: (i) first, we optimize the nominal open-loop trajectory of the system using a `blackbox simulation' model, (ii) then we identify the linear system governing perturbations from the nominal trajectory using random input-output perturbation data, and design an LQR controller for this linearized system.  We have shown that the performance of D2C algorithm is approximately optimal, in the sense that the decoupled design is near optimal to second order in a suitably defined noise parameter \cite{D2C1.0}. In this work, we apply the D2C approach to the optimal control of the evolution of material micro-structures governed by the phase field model.
For comparison, we consider RL techniques \cite{sutton2018reinforcement}. The past several years have seen significant progress in deep neural networks based reinforcement learning approaches for controlling unknown dynamical systems, with applications in many areas like playing games \cite{silver2016mastering}, locomotion \cite{lillicrap2015continuous} and robotic hand manipulation \cite{levine2016end}. A number of new algorithms that show promising performance are proposed \cite{acktr}\cite{trpo}\cite{ppo} and various improvements and innovations have been continuously developed. However, despite excellent performance on a number of tasks, RL is highly data intensive. The training time for such algorithms is typically very large, and high variance and reproducibility issues mar the performance \cite{henderson2018deep}. Thus, materials microstructure control is intractable for current RL techniques. 
Recent work in RL has also focused on PDE control using techniques like the Deep Deterministic Policy Gradient(DDPG) with modifications such as action descriptors to limit the large action spaces in infinite dimensional systems \cite{pan2018reinforcement}, as such algorithms tend to degrade in performance with an increase in dimensionality of the action-space. However, such an approach, in general, is not feasible for the Microstructure problem.  \\

\begin{figure}
    \centering
\begin{multicols}{2}
   
      \subfloat[Initial State]{\includegraphics[width=0.9\linewidth, height=101pt]{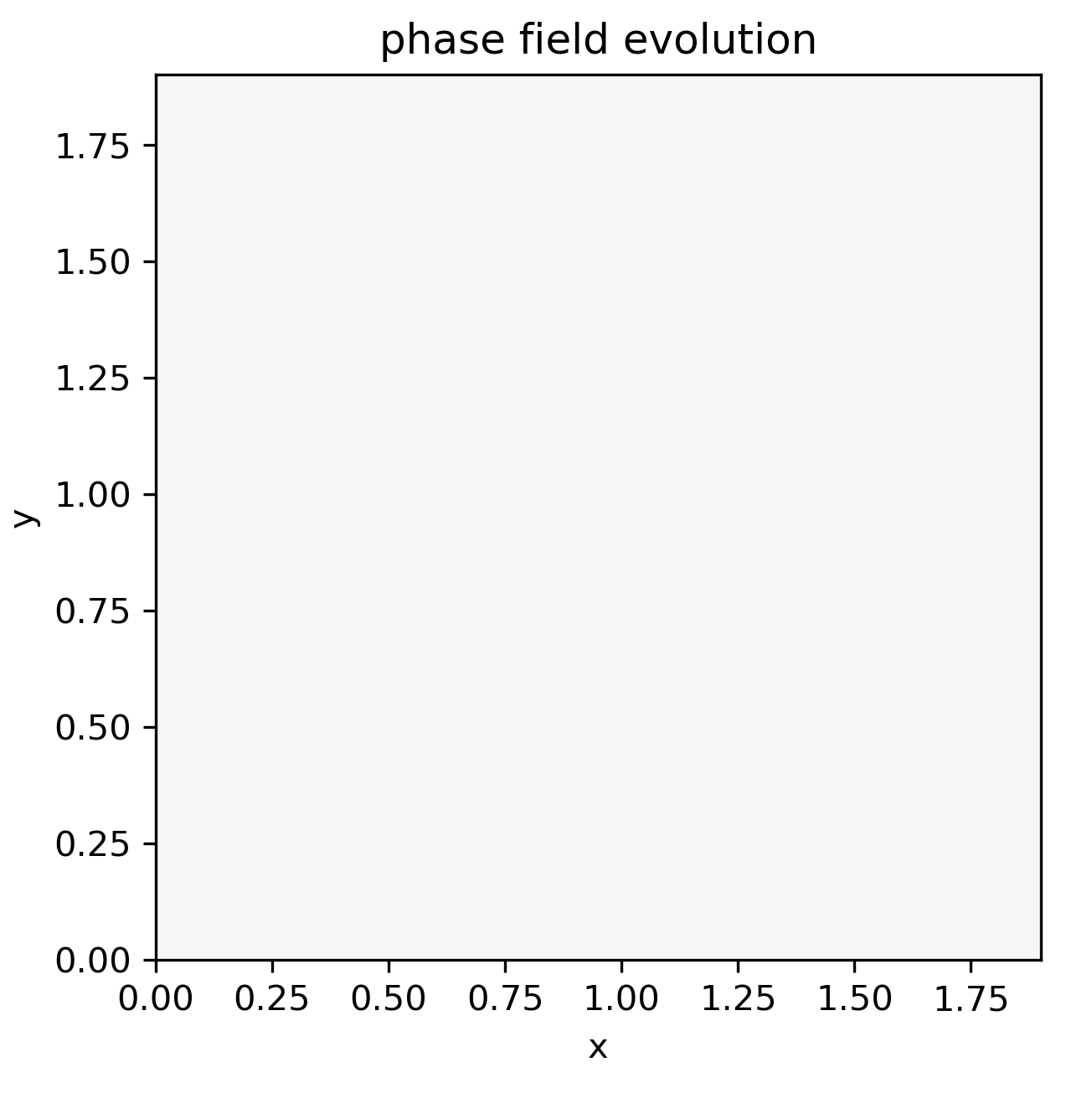}} 
      \subfloat[Goal state]{\includegraphics[width=\linewidth]{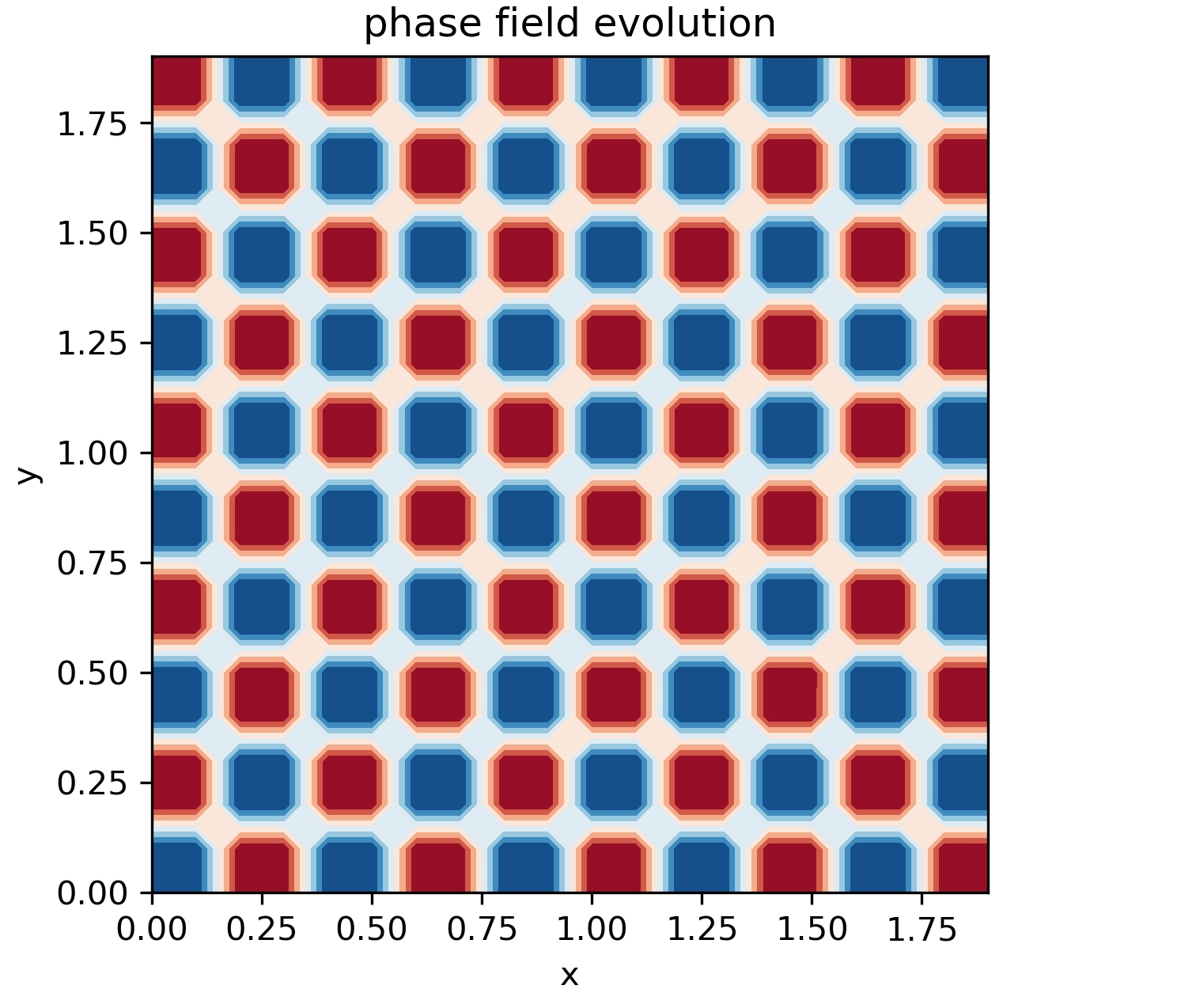}}
      
      %\subfloat[Gradient-descent based D2C Convergence]{\includegraphics[width=1\linewidth]{images/episodic_cost_OL_training.png}}
      
\end{multicols}
\begin{multicols}{2}

      \subfloat[D2C Final State]{\includegraphics[width=0.9\linewidth]{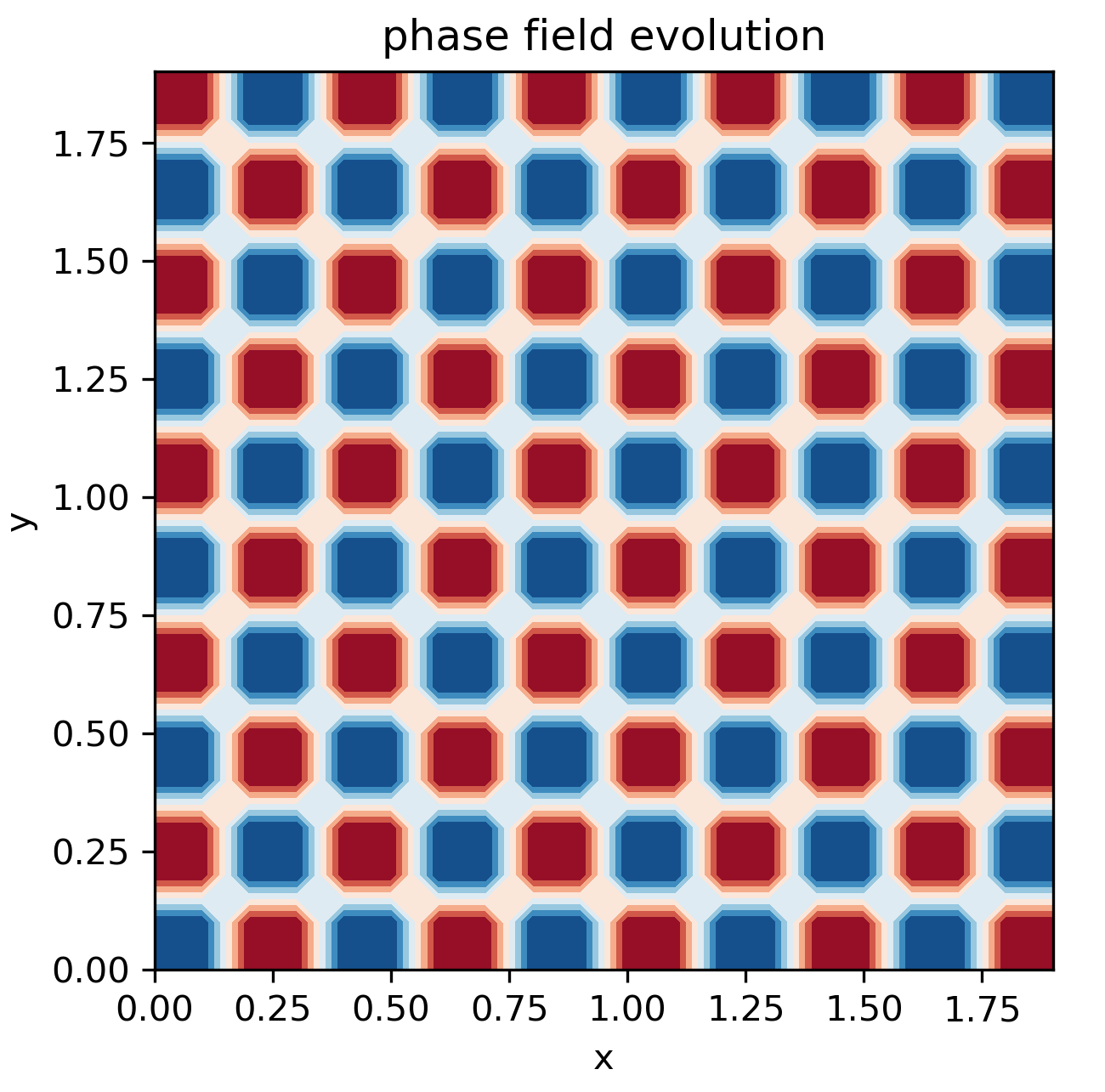}}
      \subfloat[DDPG Final State]{\includegraphics[width=0.9\linewidth]{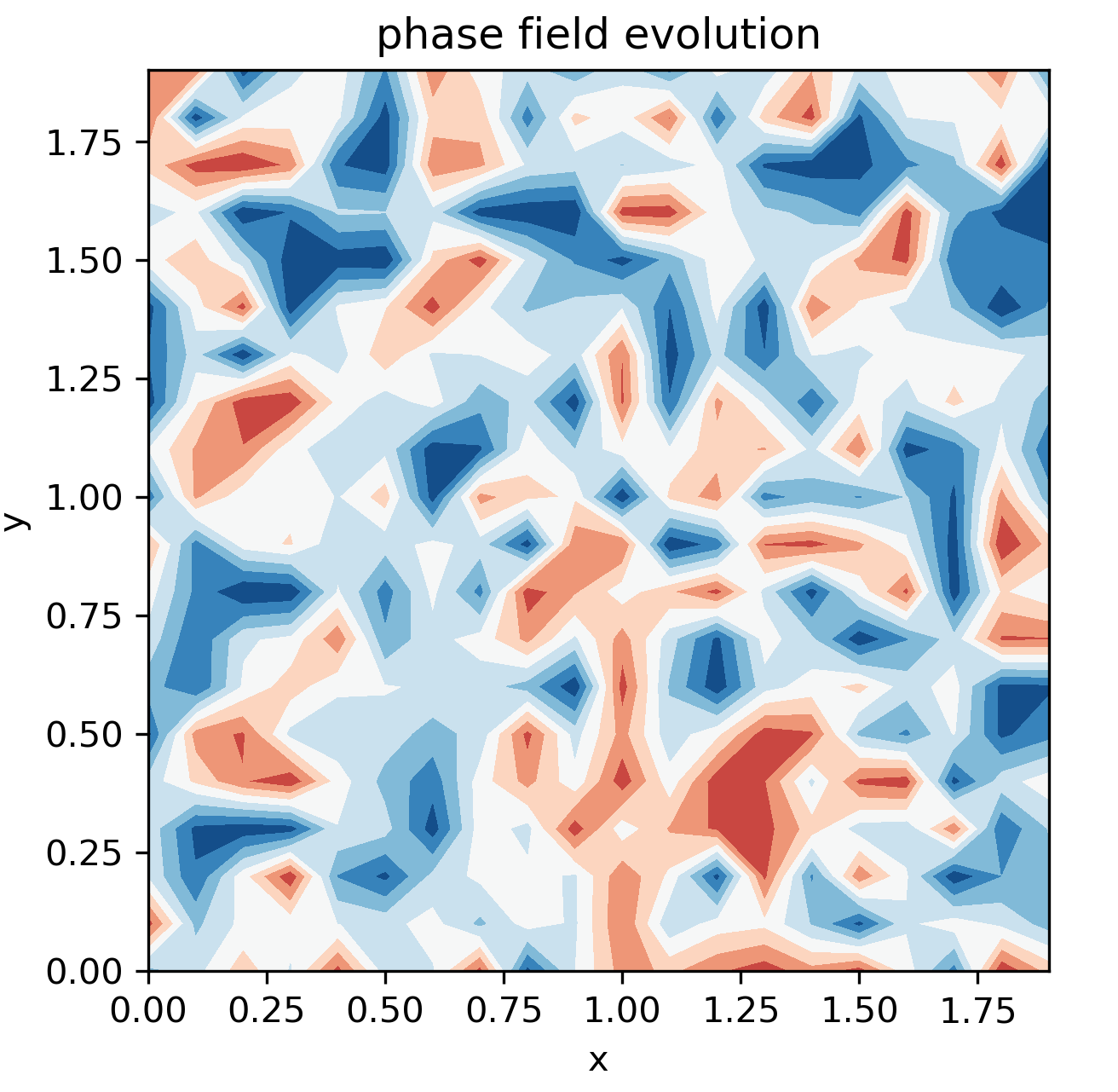}}
      %\subfloat[Gradient-descent based D2C Convergence]{\includegraphics[width=1\linewidth]{images/episodic_cost_OL_training.png}}
      
\end{multicols}

    %\caption{Comparing control cost at different time-discretizations}
    \caption{\small Material Micro-structure Control: the goal is to take the micro-structure from the initial to the goal state (top row). The micro-structure evolution is governed by a nonlinear PDE which makes its control computationally intractable. A comparison between a state of the art data based RL technique is shown versus the proposed D2C technique (bottom row) for the reliable and tractable solution of this control problem.  }
    \label{fig:up_top}
\end{figure}

%\textcolor{red}{Put a picture here that shows the superiority of using D2C to DDPG, and a sample application.}

The control design of systems governed by PDEs are known to be computationally intractable \cite{bensoussan1992representation}. Owing to the infinite dimensionality of the system, typically model reduction techniques are used to reduce the size of the state space. The most widely used method is the Proper Orthogonal Decomposition (POD)
method that finds a reduced basis for the approximation of the nonlinear PDE using a simulation of the same \cite{lall2002}. However, in general, the basis eigenfunctions can be quite
different around the optimal trajectory when compared to an initial trajectory, and an approximation using the latter’s reduced basis can lead to unacceptable performance \cite{ravindran2002},\cite{kunisch2008}. 
In our approach, the nominal optimization problem is directly solved which does not need any reduced order modeling. Furthermore, the feedback design is
accomplished by identifying a linear time-varying (LTV) model which automatically produces a reduced order model (ROM) of the system based purely on the input output data
while facilitating the use of linear control design techniques such as LQG/ILQR on the problem as opposed to having to solve a nonlinear control problem in the methods stated previously. Furthermore, the entire control design for the problem is done purely in a data based fashion since the technique only queries a black box simulation model of the process. \\

The contributions of the paper are as follows: we show how to model the dynamics of a multi-phase micro-structure and unveil its underlying structure. We present the application of the D2C approach, and a state of the art RL technique (DDPG) to the control of such microstructure dynamics for the first time in the literature, to the best of our knowledge. Our results show that the local D2C approach outperforms the global RL approach such as DDPG when operating on higher dimensional phase field PDEs. We also show that the D2C approach is robust to noise in the practical regimes, and that global optimality can be recovered in higher noise levels through an open-loop replanning approach. Furthermore, we can exploit the peculiar physical properties of material micro-structure dynamics to optimize the control actuation architecture that leads to highly computationally efficient control that does not sacrifice much performance.
We envisage that this work is a first step towards the construction of a systematic feedback control synthesis approach for the control of material micro-structures, potentially scalable to real applications.\\

%\section{Related Work}
The rest of the paper is organized as follows: In Section \ref{dynamics}, the micro-structure dynamics are expanded upon, describing the different PDEs tested. We propose the decoupled data based control (D2C) approach in Section \ref{sec_alg}. In Section \ref{secsim}, the proposed approach is illustrated through custom-defined test problems of varying dimensions, along with testing robustness to noise, followed by improvements to the algorithms which exploit the physics of the system. 

\section{Microstructure Dynamics: Classes of PDEs Implemented}
\label{dynamics}
This section provides a brief overview of the non-linear dynamics model of a general multi-phase micro-structure. We assume an infinitely large, 2-dimensional structure satisfying \textit{periodical boundary} conditions. 

\subsection{The material system}
The evolution of material micro-structures can be represented by two types of partial differential equations, i.e., the Allen-Cahn equation \cite{ALLEN19791085} representing the evolution of a  non-conserved quantity, and the Cahn-Hilliard equation \cite{Cahn1958} representing the evolution of a conserved quantity. The Allen-Cahn equation has a general form of  
\begin{align}
    &\frac{\partial \phi}{\partial t}= -M(\frac{\partial F}{\partial\phi}-\gamma\nabla^{2}\phi) \label{eq:Dyn_main_ac}
\end{align}
while the Cahn-Hilliard equation has the form
\begin{align}
    &\frac{\partial \phi}{\partial t}= \nabla \cdot M \nabla (\frac{\partial F}{\partial\phi}-\gamma\nabla^{2}\phi) \label{eq:Dyn_main_ch}
\end{align}
where $\phi=\phi(x,t)$ is called the ‘order parameter’, which is a spatially varying quantity. In Controls parlance, $\phi$ is the state of the system, and is infinite dimensional, i.e., a spatio-temporally varying function. It reflects the component proportion of each phase of material system. For a two-phase system studied in this work, $\phi$ = -1 represents one pure phase and $\phi$ = 1 represent the other, while $\phi$ $\in$ (-1, 1) stands for a combination state of both phases on the boundary interface between two pure phases; \textit{M} is a parameter related to the mobility of material, which is assumed \textit{constant} in this study; \textit{F} is an energy function with a non-linear dependence on $\phi$; $\gamma$ is a gradient coefficient controlling the diffusion level or thickness of the boundary interface.

% Eq. \ref{eq:Dyn_main} is solved using the finite-difference method, a detailed account of which is provided by Fleck \textit{et al}. \cite{Fleck2011}, and serves as the governing equation for phase field models of the two-phase microstructure system. 

\begin{figure}[!htb]
\centering
     \includegraphics[width=\linewidth]{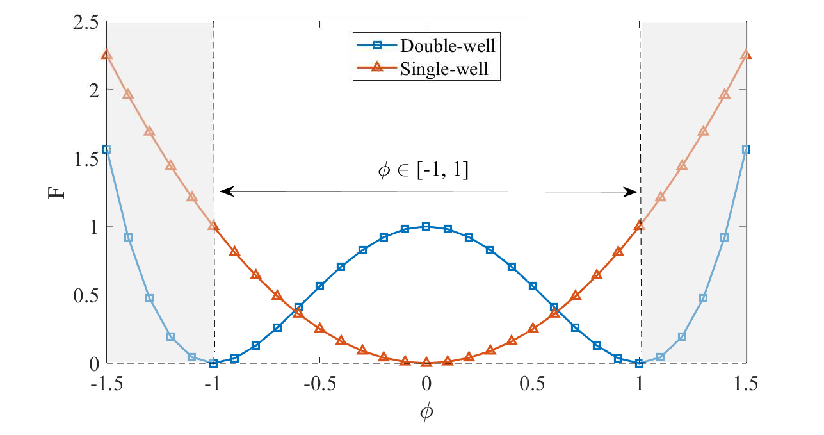}
\caption{Illustration of double-well and single-well potential function F.}
\label{F}
\end{figure}

In essence, the Phase Field Method is one of gradient flow methods, which means the evolution process follows the path of steepest descent in an energy landscape starting from an initial state until arriving at a a local minimum. So, the behavior of micro-structures in phase field modeling highly depends on the selection of energy density function \textit{F}. For instance, the double-well potential function owing to two minima at $\phi$ = $\pm$ 1  and separated by a maximum at $\phi$ = 0 (as plotted in Fig. \ref{F})  will cause the field $\phi$ to separate into regions where $\phi \approx \pm$ 1, divided by the boundary interface, while the single-well potential function owing to a single minimum at $\phi$ = 0 (see Fig. \ref{F}) predicts a gradual smoothing of any initial non-uniform $\phi$ field, yielding a uniform field with $\phi$ = 0 in the long-time limit. 

Accordingly, the evolution of material micro-structure can be governed by selecting proper form of energy density function F. Controlling the evolution process of micro-structures represented by Allen-Cahn equation and Cahn-Hilliard diffusion equation are completely distinct. The former one is governed through generating or deleting order parameter straightforwardly, while the latter is done through guiding the transport and redistribution of the conserved order parameter across the whole domain. 

In this study, we adopt the following general form of energy density function \textit{F}: %\textcolor{blue}{Please give reference}:
\begin{align}
    &F(\phi;T,h) = \phi^{4}+T\phi^{2}+h\phi \label{eq:D_well} 
\end{align}
Here, the parameters \textit{T} and \textit{h} are not governed by the A-C and C-H equations. Instead, they may be externally set to any value and may be spatially as well as temporally varying. 
%\textit{
%}

\subsection{Discretized model for PDE control}
The phase-field is discretized into a 2D grid of dimension $N\times N$. Let $\phi^{i,j}_t$ denote the phase field of the cell defined by grid location $(i,j)$ for all $i,j=\{1,...,N\}$, at time $t$.

From Eq.~(\ref{eq:Dyn_main_ac}) and (\ref{eq:D_well}), we have the A-C Equation of the form

\begin{align}
    &\frac{\partial \phi}{\partial t}= -M(4\phi^3+2T\phi+h-\gamma\nabla^{2}\phi) \label{eq:AC}
\end{align}

Now, we apply the above discretization to Eq.~(\ref{eq:AC}), and use a central-difference scheme to get the high-dimensional 2D phase-field model. Re-arranging the terms gives the time-variation of the local phase-field:

\begin{equation}
\begin{split}
    \phi^{i,j}_{t+1}=\phi^{i,j}_t -M\Delta t \{4(\phi^{i,j}_t)^3 + 2T^{i,j}_t\phi^{i,j}_t+ h^{i,j}_t- \\\gamma (\frac{\phi^{i+1,j}_t+\phi^{i-1,j}_t+\phi^{i,j+1}_t+\phi^{i,j-1}_t-4\phi^{i,j}_t}{\Delta x^2})\}
\end{split}
\label{eq:AC-CD}
\end{equation}
We can further separate the state and control terms to get
\begin{equation}
\begin{split}
    \phi^{i,j}_{t+1}=\phi^{i,j}_t -M\Delta t \{4(\phi^{i,j}_t)^3 +  \\\gamma (\frac{\phi^{i+1,j}_t+\phi^{i-1,j}_t+\phi^{i,j+1}_t+\phi^{i,j-1}_t-4\phi^{i,j}_t}{\Delta x^2})\} \\- \{M\Delta t \begin{bmatrix}2\phi^{i,j}_t &1\end{bmatrix} u^{i,j}_t\} 
\end{split}
\label{eq:AC-CD}
\end{equation}
with the periodic boundary conditions $\phi^{i,N+1}_t=\phi^{i,1}_t$ and $\phi^{N+1,j}_t=\phi^{1,j}_t$. 

Similarly, from Eq.~(\ref{eq:Dyn_main_ch}), we can write a central-difference scheme for the C-H Equation. 

\begin{equation}
\begin{split}
    \phi^{i,j}_{t+1}=\phi^{i,j}_t -M\Delta t \{\\(\frac{\mu^{i+1,j}_t+\mu^{i-1,j}_t+\mu^{i,j+1}_t+\mu^{i,j-1}_t-4\mu^{i,j}_t}{\Delta x^2})\}
\end{split}
\label{eq:CH-CD}
\end{equation}
where $\mu^{i,j}_t = -\{4(\phi^{i,j}_t)^3+2T^{i,j}_t\phi^{i,j}_t+h^{i,j}_t-\gamma (\frac{\phi^{i+1,j}_t+\phi^{i-1,j}_t+\phi^{i,j+1}_t+\phi^{i,j-1}_t-4\phi^{i,j}_t}{\Delta x^2})\}$.

Given the phase field vector $\phi_t$ and the control input vector $u_t$ at any time $t$, as a stack of all states $\{\phi^{i,j}_t\}$ and control inputs $\{u^{i,j}_t\}=\{T^{i,j}_t, h^{i,j}_t\}^T$ respectively, we can simulate the phase-field at the next time step from Eq.~(\ref{eq:AC-CD}) and (\ref{eq:CH-CD}). We also observe the dynamics to be affine in control for both cases.

The system can, thus, be described in the discretized state-space form as 

\begin{align}
    &\Phi_{t+1}=f(\Phi_t) + g(\Phi_t)U_t, \label{eq:disc_time_eq}
\end{align}
%&\frac{\phi_{t+1}-\phi_t}{\delta t}=-M(4\phi_t^3 + 2T_t\phi_t + h_t)
where $\Phi_t = \{\phi_t^{i,j}\}$ is the phase variable on the spatially discretized FD grid at time $t$, and $U_t = \{u_t^{i,j}\}$ represents the control variables on the same grid at time $t$.

\subsection{Minimization of energy function $F$ for controlling micro-structure dynamics}
\begin{figure}[!htb]

\begin{multicols}{2}
    \hspace{0cm} 
      \subfloat[Effect of T on F]{\includegraphics[width=1.0\linewidth]{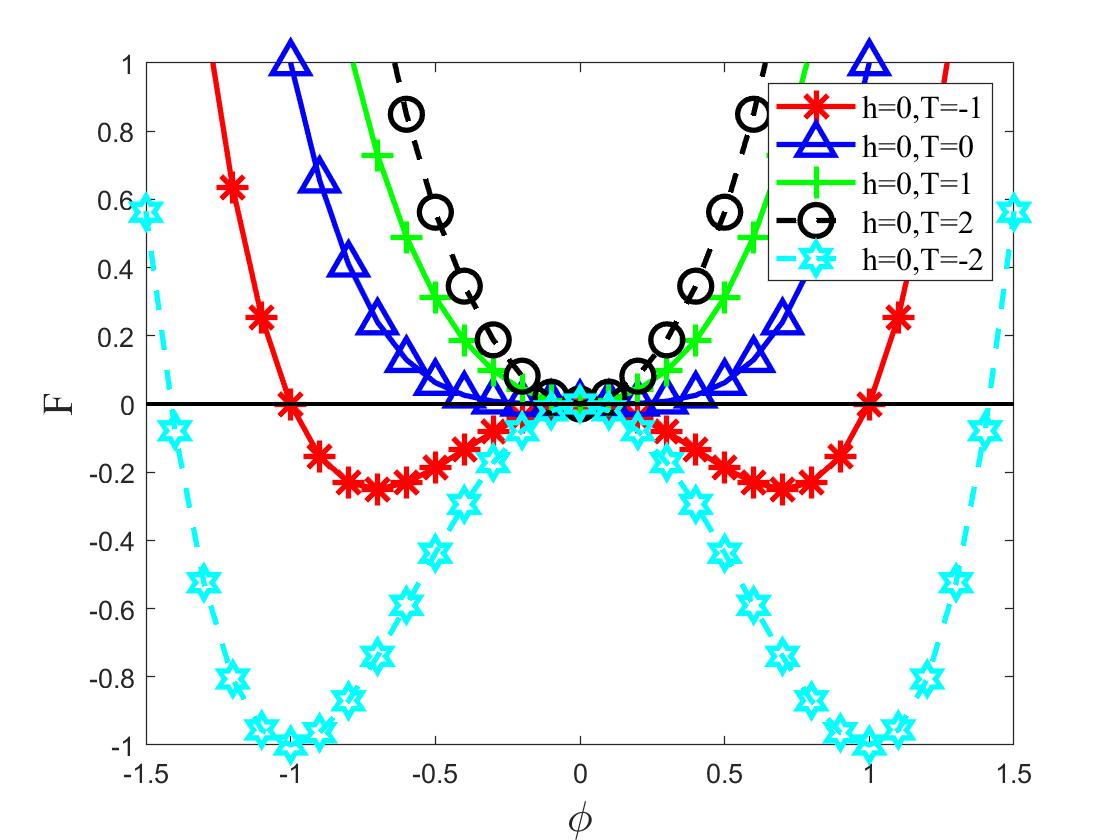}}    
      \subfloat[Effect of h on  F]{\includegraphics[width=\linewidth]{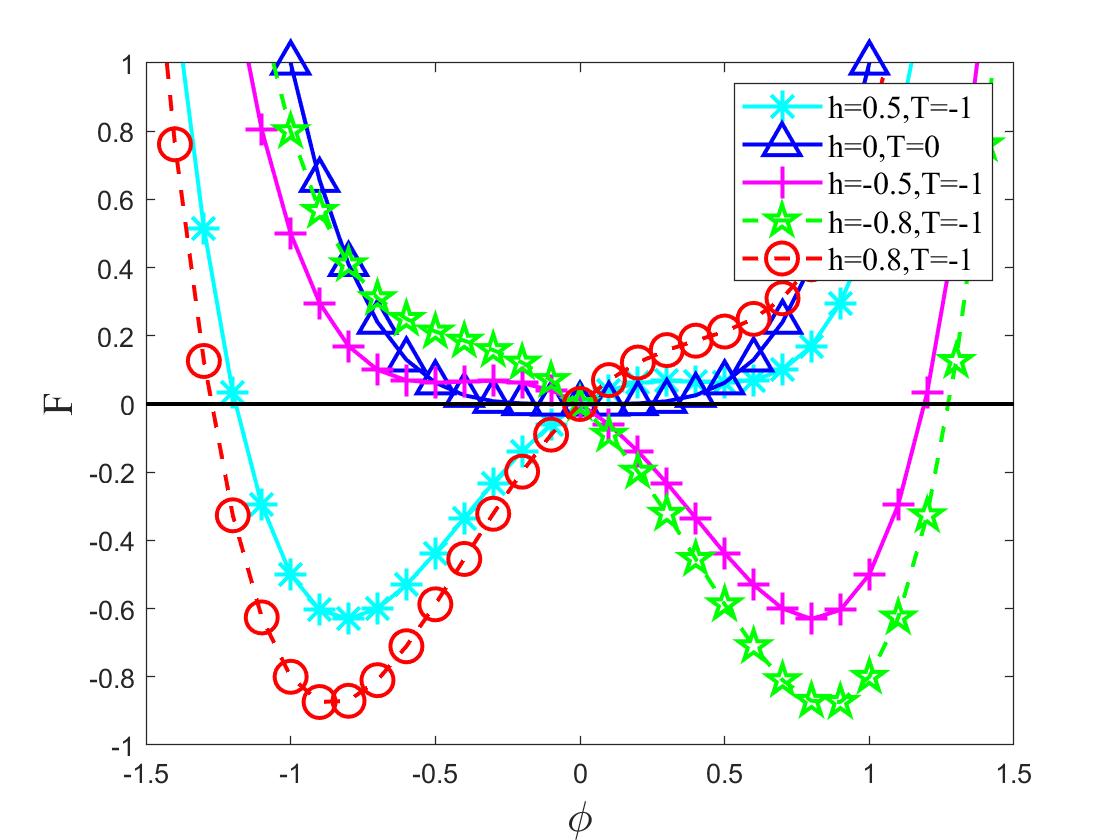}}
\end{multicols}
\caption{{Effect of T and h on energy function F}}
\label{1}
\end{figure}

The control variables $T$ and $h$ govern the evolution process of micro-structure by influencing the slope and minimums of energy density function $F$ as shown in Fig. \ref{1}. Suppose that $h$ = 0, a positive $T$ yields a double-well $F$ while negative $T$ gives a single-well $F$, as seen in Fig. \ref{1}(a). As explained earlier, under $h$ = 0, $T$ determines the $\Phi$ phase will go to a uniform state or phase-separating state in the long-time limit. When $h \neq$ 0, the symmetric energy profile shown in Fig. \ref{1}(a) becomes skewed as shown in Fig. \ref{1}(b). Under this condition, the parameter $h$ determines which phase will be favored during evolution. If $h>0$, the minimum locates on the right hand side, and therefore the phase represented by $\Phi=1$ is favored while, if $h<0$, then the $\Phi$ field favors the phase $\Phi=-1$. 
% Understanding the properties of the energy density function {$F$} in Eq. \ref{eq:D_well} is crucial to achieve its minimization and to realize the control of the Allen-Cahn equation. 

In most applications, \textit{T} and \textit{h}  represent specific input conditions, \textit{i.e.}, representing temperature and an external force field (\textit{e.g}. electromagnetic or mechanical), respectively. Since we aim at investigating the control of material micro-structure evolution during processing, we suppose that \textit{T} and \textit{h} can be adjusted to any value instantaneously, and that the feedback to the controller is the entire phase field $\Phi$ (please see the Simulations section for details).

% \subsection{Difference of Allen-Cahn and Cahn-Hilliard control}
% Determined by the different orders of derivatives of control variables ($T$ and $h$) in the evolution equations, the evolving behaviors of order parameter field from initial state to the desired goal state governed by Allen-Cahn and Cahn-Hilliard equations are completely distinct. For the Allen-Cahn equation, $T$ and $h$ are taken no derivative, and the value of order parameter can be therefore changed straightforwardly by choosing proper $T$ and $h$ (see $a_0-a_4$ in Fig. \ref{ACCH}). While for the Cahn-Hilliard equation, a Laplacian operation is applied to the term containing $T$ and $h$, which causes the order parameter field evolves to the desired state through the transport of order parameter field (see $b_0-b_4$ in Fig. \ref{ACCH}). Consequently, the demanded cost of controlling Allen-Cahn model to evolve from a given initial state to a goal state depends on the discrepancy of the initial state from the desired goal state, while that for Cahn-Hilliard model depends on the distribution of order parameter field as well.

% \begin{figure*}[!htb]
%     \centering
%     \includegraphics[width=1\linewidth]{images/AC_CH.png}
%     \caption{Evolution process of order parameter field governed by Allen-Cahn ($a_0-a_4$) and Cahn-Hilliard ($b_0-b_4$) equations. $a_0$ and $b_0$ represent the initial states that be controlled to evolve to the final states represented by $a_4$ and $b_4$, respectively.}
%     \label{ACCH}
% \end{figure*}

%#LOREM IPSUM RANDOM CRAP TO BE REPLACED ASAP

%\input{MaterialControl.tex} 
\section{Decoupled Data Based Control (D2C) Algorithm}\label{sec_alg}

In this section, we propose a data-based approach to solve the material phase-separation control problem. %as an alternative to the model-based approach in the previous section. 
In particular, we apply the so-called decoupled data-based control (D2C) algorithm that we recently proposed \cite{D2C1.0}, and extend it for high dimensional applications. The D2C algorithm is a highly data-efficient Reinforcement Learning (RL) method that has shown to be much superior to state of the art RL algorithms such as the Deep Deterministic Policy Gradient (DDPG) in terms of data efficiency and training stability while retaining similar or better performance. In the following, we give a very brief introduction to the D2C algorithm (please see \cite{D2C1.0, cdc_soc, aistatsd2c2} for the relevant details). 

Let us reconsider the dynamics of the system given in Eq.~(\ref{eq:Dyn_main_ac}) and (\ref{eq:Dyn_main_ch}) and rewrite it in a discrete time, noise perturbed state space form as follows:
\begin{equation}
\Phi_{t+1} = f(\Phi_t) + g(\Phi_t)(U_t + \epsilon w_t),
\end{equation}
where $\Phi_t$ is the state, $U_t$ is the control, $w_t$ is a white noise perturbation to the system, and $\epsilon$ is a small parameter that modulates the noise in the system. Suppose now that we have a finite horizon objective function:
$ 
J(\Phi_0) = E[\sum_{t=0}^{T-1}c(\Phi_t,U_t) + \Psi(\Phi_T)],
$ where $c(\Phi,u)$ denotes a running incremental cost, $\Psi(\Phi)$ denotes a terminal cost function and $E[\cdot]$ is an expectation operator taken over the sample paths of the system. The objective of the control design is then to design a feedback policy $U_t(\Phi_t)$ that minimizes the cost function above and is given by $J^*(\Phi_0) = \min_{U_t(.)}E[\sum_{t=0}^{T-1}c(\Phi_t,U_t) + \Psi(\Phi_T)]$. 
% \begin{equation}
%     J^*(x_0) = \min_{u_t(.)}E[\sum_{t=0}^{T-1}c(x_t,u_t) + \phi(x_T)].
% \end{equation}

The D2C algorithm then proposes a 3 step procedure to approximate the solution to the above problem.

First, a noiseless open-loop optimization problem is solved to find an optimal control sequence, $\bar{U}^*_t$, that generates the nominal state trajectory $\bar{\Phi}^*_t$:
\begin{equation} \label{D2C-OL}
    J^*(\Phi_0) = \min_{U_t}(\sum_{t=0}^{T-1}c(\Phi_t,U_t) + \Psi(\Phi_T)).
\end{equation}
subject to the zero noise nominal dynamics: $\Phi_{t+1} = f(\Phi_t)+g(\Phi_t)U_t$.

Second, the perturbed time varying linear system around the nominal given by $\delta \Phi_{t+1} = A_t\delta \Phi_t + B_t \delta U_t$ is identified in terms of the time varying system matrices $A_t$, $B_t$.

Third, an LQR controller for the above time varying linear system is designed whose time varying gain is given by $K_t$. Finally, the control applied to the system is given by $U_t(\Phi_t) = \bar{U}^*_t + K_t \delta \Phi_t$.

% \textit{Near Optimality of D2C:} Let the mean and variance of the true cost function be given by $\bar{J}^*$ and $Var[J]^*$ respectively. Then, it can be shown that the  D2C procedure outlined above is near optimal to the second order in the small parameter $\epsilon$ in the sense that $|\bar{J} - \bar{J}^*|$ is $O(\epsilon^2)$ and moreover, $\sqrt{|Var[J] - Var[J]^*|}= O(\epsilon^2)$, where $\bar{J}$ and $Var[J]$ are the mean and variance of the cost corresponding to the D2C policy above (refer to \cite{D2C1.0} for the details).  

The following subsections provide details for each of the above-mentioned steps of the D2C algorithm.

\subsection{Open Loop Trajectory Optimization}
\label{sec_open_ilqr}
We utilize an ILQR based method to solve the open-loop optimization problem. ILQR typically requires the availability of analytical system Jacobian, and thus, cannot be directly applied when such analytical gradient information is unavailable (much like Nonlinear Programming software whose efficiency depends on the availability of analytical gradients and Hessians).  In order to make it an (analytical) model-free algorithm, it is sufficient to obtain estimates of the system Jacobians from simulations, and a sample-efficient randomized way of doing so is described in the following subsection. Since ILQR is a well-established framework, we skip the details and instead present pseudocode in algorithm \ref{model_free_DDP_OL}.

\subsubsection{Estimation of Jacobians: Linear Least Squares by Central Difference (LLS-CD)}
\label{sys_id_solve}

Using Taylor's expansions of the dynamics `$h$',  where $h (\Phi,U) = f(\Phi) + g(\Phi)U$ is the non-linear model of Section 2), about the nominal trajectory $(\bar{\Phi}_t,  \bar{U}_t)$ on both the positive and the negative sides, we obtain the following central difference equation:
$
    h(\bar{\Phi}_t + \delta \Phi_t, \bar{U}_t + \delta U_t) - h(\bar{\Phi}_t - \delta \Phi_t, \bar{U}_t - \delta  U_t) \\ = 2 \begin{bmatrix} h_{\Phi_t} & h_{U_t} \end{bmatrix} \begin{bmatrix}  \delta {\Phi_t} \\ \delta {U_t} \end{bmatrix} + O(\| \delta {\Phi_t}\|^3 + \| \delta {U_t}\|^3).
$
Multiplying by $\begin{bmatrix} \delta {\Phi_t}^T & \delta {U_t}^T \end{bmatrix}$ on both sides to the above equation and apply standard Least Square method:
\begin{equation}
    \begin{split}
    &\begin{bmatrix} h_{\Phi_t}~h_{U_t} \end{bmatrix} = H \delta Y_t^T (\delta Y_t\delta Y_t^T)^{-1} \\
    &H = \begin{bmatrix} h({\bar{\Phi}_t} + \delta {\Phi_t^1}, {\bar{U}_t} + \delta {U_t^1}) - h({ \bar{\Phi}_t} - \delta {\Phi_t^1}, {\bar{U}_t} - \delta {U_t^1})\\ h({\bar{\Phi}_t} + \delta {\Phi_t^2}, {\bar{U}_t} + \delta {U_t^2}) - h({\bar{\Phi}_t} - \delta {\Phi_t^2}, {\bar{U}_t} - \delta {U_t^2}) \\ \vdots  \\h({\bar{\Phi}_t} + \delta {\Phi_t^{n_s}}, {\bar{U}_t} + \delta {U_t^{n_s}}) - h({\bar{\Phi}_t} - \delta {\Phi_t^{n_s}}, {\bar{U}_t} - \delta {U_t^{n_s}})\end{bmatrix} \nonumber
     \end{split}
\end{equation}
where `$n_s$' be the number of samples for each of the random variables, $\delta {\Phi_t}$ and $\delta {U_t}$. Denote the random samples as $\delta {X_t} = \begin{bmatrix} \delta {\Phi_t^1}& \delta {\Phi_t^2}& \ldots &\delta {\Phi_t^{n_s}}\end{bmatrix}$, $\delta {\mathbb{U}_t} = \begin{bmatrix} \delta {U_t^1} &\delta {U_t^2}& \ldots& \delta {U_t^{n_s}}\end{bmatrix}$ and $\delta Y_t = \begin{bmatrix} \delta X_t & \delta \mathbb{U}_t \end{bmatrix}$.

We are free to choose the distribution of $\delta {\Phi_t}$ and $\delta {U_t}$. We assume both are i.i.d. Gaussian distributed random variables with zero mean and a standard deviation of $\sigma$.~This ensures that $\delta Y_t \delta Y_t^T$ is invertible.

Let us consider the terms in the matrix $\delta Y_t \delta Y_t^T=\begin{bmatrix}  \delta {X_t} \delta {X_t}^T & \delta {X_t} \delta {\mathbb{U}_t}^T \\ \delta {\mathbb{U}_t} \delta {X_t}^T  & \delta {\mathbb{U}_t} \delta {\mathbb{U}_t}^T \end{bmatrix}$.~$\delta {X_t} \delta {X_t}^T = \sum_{i=1}^{n_s} \delta {\Phi_t}^i {\delta {\Phi_t}^i}^T$. Similarly, $\delta {\mathbb{U}_t} \delta {\mathbb{U}_t}^T = \sum_{i=1}^{n_s} \delta {U_t}^i {\delta {U_t}^i}^T$, $\delta {\mathbb{U}_t} \delta {X_t}^T = \sum_{i=1}^{n_s} \delta {U_t}^i {\delta {\Phi_t}^i}^T$ and $\delta {X_t} \delta {\mathbb{U}_t}^T = \sum_{i=1}^{n_s} \delta {\Phi_t}^i {\delta {U_t}^i}^T$. From the definition of sample variance, for a large enough $n_s$, we can write the above matrix as 
\begin{equation*}
\begin{split}
    \delta Y_t \delta Y_t^T &= \begin{bmatrix} \sum_{i=1}^{n_s} \delta {\Phi_t}^i {\delta {\Phi_t}^i}^T & \sum_{i=1}^{n_s} \delta {\Phi_t}^i {\delta {U_t}^i}^T \\ \sum_{i=1}^{n_s} \delta {U_t}^i {\delta {\Phi_t}^i}^T & \sum_{i=1}^{n_s} \delta {U_t}^i {\delta {U_t}^i}^T
    \end{bmatrix}\\ &\approx \begin{bmatrix} \sigma^2(n_s - 1) {\text I_{n_\Phi}} & {\text 0_{n_\Phi \times n_U}} \\ 0_{n_U \times n_\Phi} & \sigma^2 (n_s - 1) {\text I_{n_U}}\end{bmatrix} \\
    &= \sigma^2 (n_s - 1){\text I}_{(n_\Phi+n_U) \times (n_\Phi+n_U)}
\end{split}
\end{equation*}

\subsection{Linear Time Varying System Identification}
\label{sec_ls}
Closed-loop control design in step 2 of D2C requires the knowledge of the linearized system parameters $A_{t}$ and $B_{t}$ for $0 \leq t \leq T-1$. Here we use the standard least square method to estimate these parameters from input-output experiment data.  

First start from the perturbed linear system about the nominal trajectory and estimate the system parameters $A_{t}$ and  $B_{t}$ from:
$%$\begin{align}
\label{est_sys}
 \delta \Phi_{t+1} = \hat{A_t} \delta \Phi_t+\hat{B_t} \delta U_t,$
% $%\end{align} 
 where $\delta \Phi_{t}^{(n)}$ is the state perturbation vector and $\delta U_{t}^{(n)}$ is the control perturbation vector we feed to the system at step $t$, $n$\textsuperscript{th} simulation. All the perturbations are zero-mean, i.i.d, Gaussian noise with covariance matrix $\sigma I$. $\sigma$ is a $o(U)$ small value selected by the user. $\delta \Phi_{t+1}^{(n)}$ denotes the deviation of the output state vector from the nominal state after propagating for one step.

Run N simulations for each step and collect the input-output data: $Y = [\hat{A_t} ~|~ \hat{B_t}] X$
% \begin{align}
% \label{sysmat}
% & Y = [\hat{A_t} ~|~ \hat{B_t}] X,
% \end{align}
and write out the components:
\begin{align}
\label{lsq1}
& Y = \begin{bmatrix}
\delta \Phi_{t+1}^{(1)}& \delta \Phi_{t+1}^{(2)} &\cdots& \delta \Phi_{t+1}^{(N)} \\
\end{bmatrix},
\nonumber\\
& X=\begin{bmatrix}
  \delta \Phi_t^{(1)} & \delta \Phi_t^{(2)}& \cdots & \delta \Phi_t^{(N)} \\
  \delta U_t^{(1)} &  \delta U_t^{(2)} & \cdots & \delta U_t^{(N)} \\
\end{bmatrix}.
\end{align}

Finally, using the standard least square method, the linearized system parameters are estimated as $[\hat{A_t} ~|~ \hat{B_t}]=YX \T(XX \T)^{-1}$.
% \begin{align}
% \label{lsq2}
% &  [\hat{A_t} ~|~ \hat{B_t}]=YX \T(XX \T)^{-1}.
% \end{align}

\subsection{Closed Loop Control Design} 
% \textcolor{red}{ditch this section, just mention that we use LQR.}
 Given the estimated perturbed linear system, we design a finite horizon, discrete time LQR \cite{LQR} along the trajectory for each time step to minimize the cost function, and use the standard Riccati equations to solve for the feedback gains.
%  $J=\delta x_T^TQ \delta x_T+\sum_{t=0}^{T-1}(\delta x_t^TQ \delta x_t+u_t^TRu_t+2 \delta x_t^TNu_t)$, subjects to $\delta x_{t+1} = \hat{A_t} \delta x_t+\hat{B_t} \delta u_t$, where $u_t=-K_t \delta x_t$. The feedback gains are calculated as $K_t=(R+B^TP_{t+1}B)^{-1}(B^TP_{t+1}A+N^T)$, where $P_t$ is solved in a back propagation fashion from the Riccati equation: $P_{t-1}=A^TP_tA-(A^TP_tB+N)(R+B^TP_tB)^{-1}(B^TP_tA+N^T)+Q, P_T=Q, N=0$. 
% The closed-loop control policy is $u_t(x_t)=\bar{u}^*_t-K_t \delta x_t$, where $\delta x_t$ is the state deviation vector from the nominal state at time step $t$.

% Given the parameter estimate of the perturbed linear system, we solve the Riccati equation for each step along the trajectory for the closed-loop optimal feedback gains $K_t$. This is a standard LQR problem, so the details are omitted here. The closed-loop control policy is $u_t(x_t)=\bar{u}^*_t+K_t \delta x_t$. Here $\delta x_t$ is the state deviation vector from the nominal state at time step $t$.

\subsection{D2C Algorithm: Summary}
The Decoupled Data Based Control  (D2C) Algorithm is summarized in Algorithm \ref{alg1}. 

\begin{algorithm}[t]
\footnotesize
  \caption{\strut Open-loop trajectory optimization via model-free ILQR}
  {\bf Input:} Initial State - ${\bf \Phi_0}$, System parameters - $\mathcal{P}$\;
  $m \gets 1$. ~~\CommentSty{/* Initialize the iteration number $m$ to 1.*/}\\
  $forward\_pass\_flag$ = true. \\
  \CommentSty{/* Run until the difference in costs between subsequent iterations is less an $\epsilon$ fraction of the former cost.*/}\\
  \While {$m==1$ {\textnormal{or}} $({\text cost}(\mathbb{T}_{nom}^{m})/{\text cost}(\mathbb{T}_{nom}^{m-1})) < 1 + \epsilon$}{
  \CommentSty{/*Each iteration has a backward pass followed by a forward pass.*/}\\
  \{$k^{m}_{0:N-1}, K^{m}_{0:N-1}$\}, $backward\_pass\_success\_flag$ $=$ Backward Pass($\mathbb{T}_{nom}^{m}$, $\mathcal{P}$).\\
  \If{backward\_pass\_success\_flag == true}{
      $\mathbb{T}_{nom}^{m+1}, forward\_pass\_flag$ $=$ Forward Pass($\mathbb{T}_{nom}^{m}$,$\{k^{m}_{0:N-1}, K^m_{0:N-1}\}$, $\mathcal{P}$).\\
      \While{forward\_pass\_flag == false}{
            $\mathbb{T}_{nom}^{m+1}, forward\_pass\_flag$ $=$ Forward Pass($\mathbb{T}_{nom}^{m}$,$\{k^{m}_{0:N-1}, K^m_{0:N-1}\}$, $\mathcal{P}$).\\
            Reduce $\alpha$ from $\mathcal{P}$.    
        }
      }
  \Else{
        Increase $\mu$ from $\mathcal{P}$.~~\CommentSty{/* Regularization step */}\\
    }
      $m \leftarrow m + 1$. \\
  $\mathbb{T}^{*}_{nom} \gets \mathbb{T}^{m+1}_{nom}$.\\
  }
  
  {\bf return $\mathbb{T}^{*}_{nom}$}\\
  \label{model_free_DDP_OL}
\end{algorithm}

\begin{algorithm}
    \caption{D2C Algorithm}
    \label{alg1}
   {\bf 1)}  Solve the deterministic open-loop optimization problem for optimal open-loop control sequence and state trajectory $(\{\bar{U}^*_t\}_{t = 0}^{T-1}, \{\bar{\Phi}_t^*\}_{t = 0}^T)$ %(Section \ref{sec_open_ilqr}). 
   using ILQR (Section \ref{sec_open_ilqr}).\\
  {\bf 2)} Linearize and identify the LTV system parameters $(\hat{A}_t, \hat{B}_t)$ via least square (Section \ref{sec_ls}).\\
  {\bf 3)} Solve the Riccati equations for each step along the nominal trajectory for feedback gain $\{ K_t \}_{t = 0}^{T-1}$.\\
  {\bf 4)} Apply the closed-loop control policy,
  
\While{$t < T$}{
  \begin{align}
& U_t = \bar{U}_t^* + K_t \delta \Phi_t, \nonumber \\
& \Phi_{t + 1} = f(\Phi_{t}) + B_{t} (U_{t} + \epsilon w_{t}),\nonumber \\
& \delta \Phi_{t+1} = \Phi_{t + 1}-\bar{\Phi}^*_{t + 1}
\end{align}
 $t = t + 1$.
}
\end{algorithm}

\section{Simulation Results}\label{secsim}

In this section, we compare the training and performance of the data based D2C approach, with the RL-based Deep Deterministic Policy Gradient(DDPG), on a material with control inputs as the temperature and external field inputs on subsets of grid points. 

\subsection{Structure and Task}
We simulated the phase-separation dynamics in Python, through calling an explicit, second-order solver subroutine in FORTRAN. The system and its tasks are defined as follows: 

% \begin{figure}[!htb]

% \begin{multicols}{2}
%     %\hspace{1.4cm} 
%       \subfloat[Initial state]{\includegraphics[width=\linewidth, height=100]{images/Initial_state.PNG}}    
%       \subfloat[Goal state-I]{\includegraphics[width=\linewidth]{images/Final_state.PNG}}

% \end{multicols}
% \begin{multicols}{2}
%     %\hspace{1.4cm} 
          
%       \subfloat[Goal state-II]{\includegraphics[width=\linewidth]{images/Final_state_2.PNG}}
%       \subfloat[Goal state-III]{\includegraphics[width=\linewidth]{images/Final_state_4.PNG}}

% \end{multicols}
% \caption{{Model simulated in Python}}
% \label{figinit}
% \end{figure}

\begin{figure}[!htb]

\begin{multicols}{3}
    %\hspace{1.4cm} 
    %   \subfloat[Initial state]{\includegraphics[width=\linewidth]{images/Initial_state.PNG}}    
      \subfloat[Goal state-I]{\includegraphics[width=\linewidth]{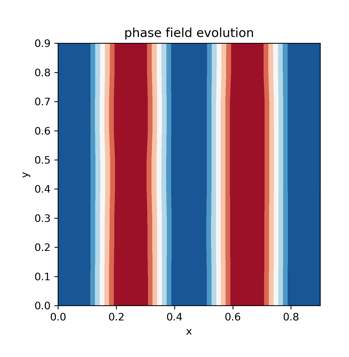}}
      \subfloat[Goal state-II]{\includegraphics[width=\linewidth]{images/Final_state_2.png}}
      \subfloat[Goal state-III]{\includegraphics[width=\linewidth]{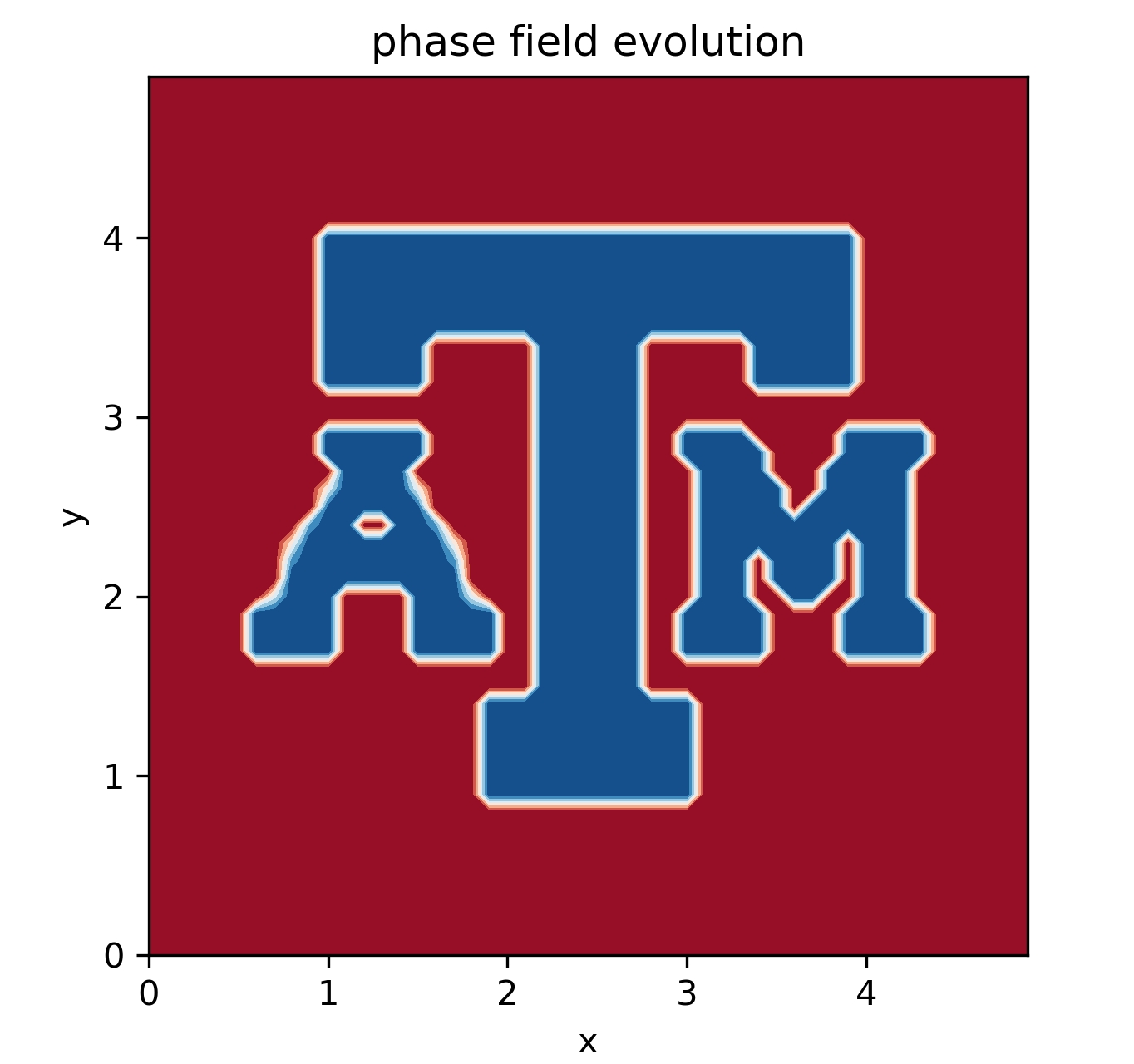}}

\end{multicols}
\caption{{Model simulated in Python}}
\label{figinit}
\end{figure}

\paragraph{Material Micro-structure}
The material model simulated consists of a 2-dimensional grid of dimensions 10x10, 20X20 and 50X50 i.e.,  100, 400 and 2500 fully observable states respectively.
%The order parameter at each of the grid point can be varied in the range $[-1, 1]$. 
Furthermore, the model is solved using an explicit, second order, central-difference based scheme. The control inputs $(T, h)$ are actuated at each grid point separately; thus, the number of control variables is twice the total number of observable states(200, 800, 5000 resp.). 
The control task is to influence the material dynamics to converge at a 
\begin{enumerate}
    \item Banded phase-distribution (\textbf{Fig.\ref{figinit}(a)}), with 100 state variables and 200 control channels
    \item Checkerboard phase-distribution (\textbf{Fig.\ref{figinit}(b)}), with 400 state variables and 800 control channels
    \item A custom phase-distribution (\textbf{Fig.\ref{figinit}(c)}), with 2500 state variables and 5000 control channels
\end{enumerate}
starting from a uniform initial condition $\phi_0=0$.

% The desired final state of the model are shown in Fig. \ref{figinit}. 

\subsection{Algorithms Tested}
% \input{Hybrid.tex}
%\subsubsection{Model Based Control}
%% REMOVE THIS LINE FOR JOURNAL
\paragraph{Deep Deterministic Policy Gradient} Deep Deterministic Policy Gradient (DDPG) is a policy-gradient based off-policy reinforcement learning algorithm that operates in continuous state and action spaces. It relies on two function approximation networks one each for the actor and the critic. The critic network estimates the $Q(s, a)$ value given the state and the action taken, while the actor network engenders a policy given the current state. Neural networks are employed to represent the networks. 
%% REMOVE THIS LINE FOR JOURNAL
%\input{MaterialControl.tex}
%\input{DDPGAlgorithm.tex}

\subsection{Training and Testing}
D2C implementation is done in three stages corresponding to those mentioned in the previous section, and a black box phase field model is simulated in Python. 
% The model based control (MBC) approach does not require any training time since it is a closed-form solution.

%\begin{figure}[!htb]
%\begin{multicols}{1}
    %\hspace{1.4cm}    
%     {\includegraphics[width=1\linewidth]{images/file49.png}}    
      
%\end{multicols}
%\caption{Model in terminal state with open loop trajectory: D2C}
%\label{figfinal}
%\end{figure}

%\begin{figure}[!htb]

%\begin{multicols}{2}
    %\hspace{1.4cm} 
%      \subfloat[Initial state]{\includegraphics[width=1\linewidth]{images/Initial_state.PNG}}    
 %     \subfloat[Goal state]{\includegraphics[width=1\linewidth]{images/file49.png}}

\begin{figure*}[!htpb]
\centering
% \begin{multicols}{3}
   
%       \subfloat[DDPG Convergence]{\includegraphics[width=1\linewidth]{images/DDPG_Plots/ddpg_banded_h_100.png}}    
%       \subfloat[Gradient-descent based D2C Convergence]{\includegraphics[width=1\linewidth]{images/Training/Banded_episodic_cost_OL_training.png}}
%       \subfloat[ILQR-based D2C Convergence]{\includegraphics[width=1\linewidth]{images/Training/Banded_D2C_2.png}}\\

% \end{multicols}
% \begin{multicols}{3}
   
%       \subfloat[DDPG Convergence]{\includegraphics[width=1\linewidth]{images/DDPG_Plots/ddpg_checkerboard_h_10_10000r.png}}    
%       \subfloat[Gradient-descent based D2C Convergence]{\includegraphics[width=1\linewidth]{images/Training/episodic_cost_OL_training_20x20.png}}
%       %\subfloat[Gradient-descent based D2C Convergence]{\includegraphics[width=1\linewidth]{images/episodic_cost_OL_training.png}}
%       \subfloat[ILQR-based D2C Convergence]{\includegraphics[width=1\linewidth]{images/Training/Figure_20X20.png}}\\

% \end{multicols}
\begin{multicols}{3}
   
      \subfloat[DDPG Convergence (Allen Cahn, GS-I)]{\includegraphics[width=1\linewidth]{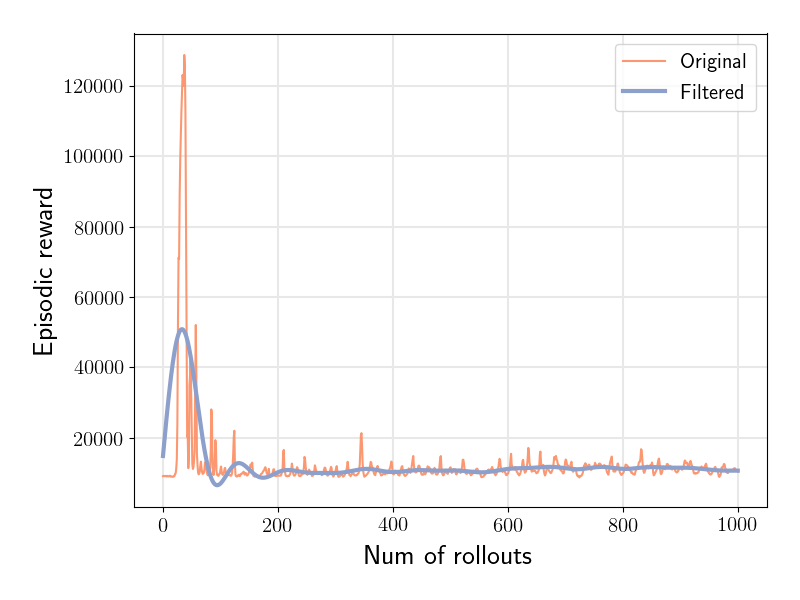}}    
      \subfloat[DDPG Convergence(Allen Cahn, GS-II)]{\includegraphics[width=1\linewidth]{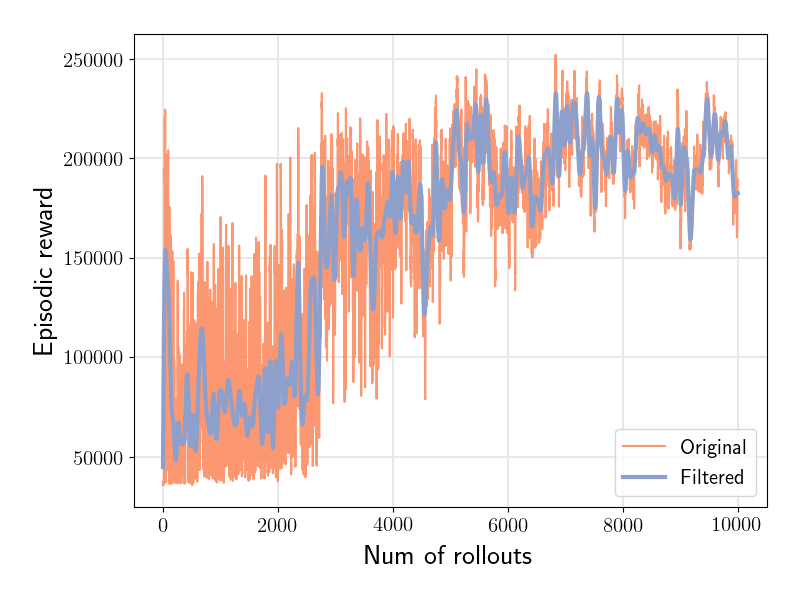}}
      %/DDPG_Plots/ddpg_checkerboard_h_10_10000r.png}}
      \subfloat[DDPG Convergence(Cahn-Hilliard, GS-I)]{\includegraphics[width=1\linewidth]{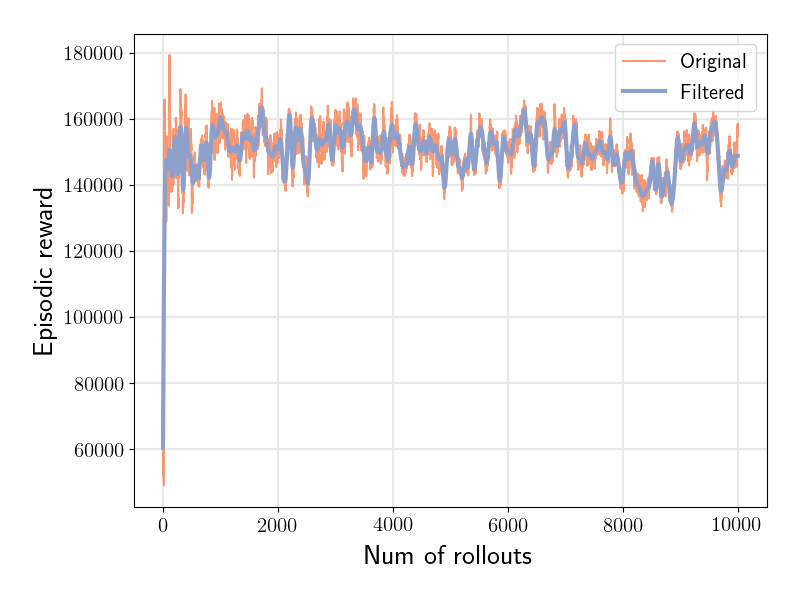}}\\

\end{multicols}
% \begin{multicols}{3}
   
%       \subfloat[D2C-GD Convergence(Allen Cahn, GS-I)]{\includegraphics[width=1\linewidth]{images/Training/Banded_episodic_cost_OL_training.png}}    
%     %   \subfloat[D2C-GD Convergence(Allen Cahn, GS-II)]{\includegraphics[width=1\linewidth]{images/Training/Figure_20X20.png}}
%       \subfloat[D2C-GD Convergence]{\includegraphics[width=1\linewidth]{images/episodic_cost_OL_training.png}}
%       \subfloat[D2C-GD Convergence(Cahn-Hilliard, GS-I)]{\includegraphics[width=1\linewidth]{images/episodic_cost_OL_training.png}}\\

% \end{multicols}

\begin{multicols}{3}
   
      \subfloat[D2C Convergence(Allen Cahn, GS-I)]{\includegraphics[width=1\linewidth]{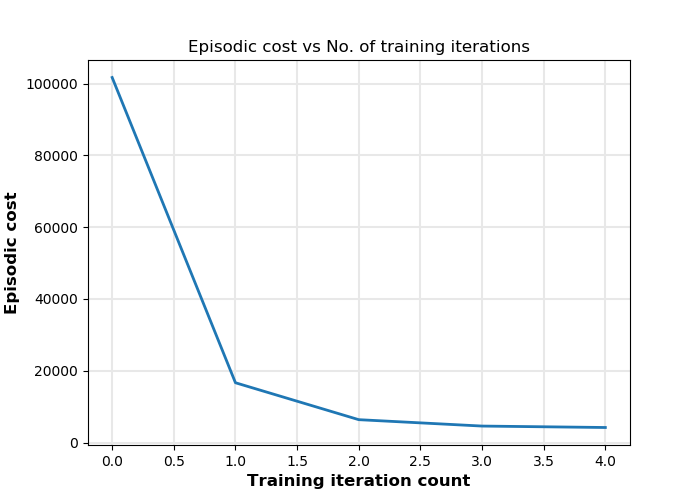}}    
      \subfloat[D2C Convergence(Allen Cahn, GS-II)]{\includegraphics[width=1\linewidth]{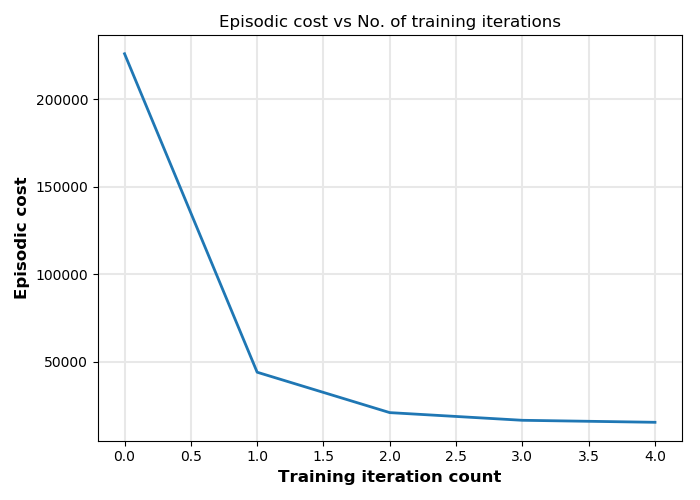}}
      %\subfloat[Gradient-descent based D2C Convergence]{\includegraphics[width=1\linewidth]{images/episodic_cost_OL_training.png}}
      \subfloat[D2C Convergence(Cahn-Hilliard, GS-I)]{\includegraphics[width=1\linewidth]{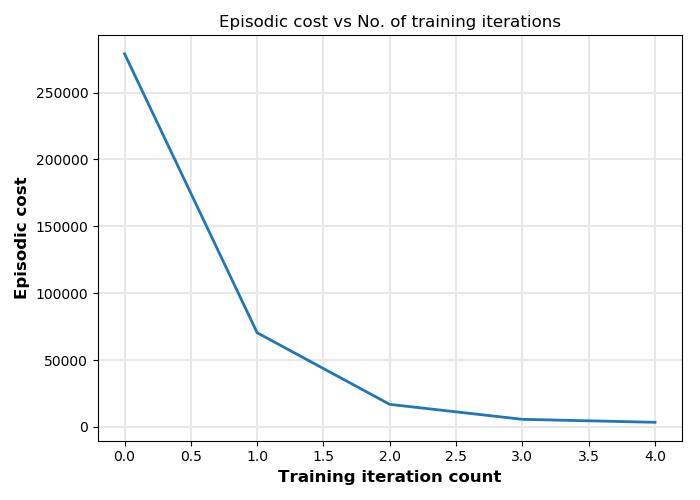}}\\

\end{multicols}

\caption{ Convergence of Episodic cost for goal-state I \textbf{(Left Column)}, goal-state II\textbf{(Middle Column)}} for Allen-Cahn PDE, and goal-state I \textbf{(Right Column)} for Cahn-Hilliard PDE. %Bottom row: Terminal state MSE during testing in D2C vs DDPG. The solid line in the plots indicates the mean and the shade indicates the standard deviation of the corresponding metric.}
\label{d2c_2_training_testing}
\end{figure*}

%\end{multicols}
%\caption{{Open loop training outcome }}
%\label{figfinal}
%\end{figure}
{\bf Training:} The open-loop training plots in Fig.~\ref{d2c_2_training_testing} show the cost curve during training. After the cost curves converge, we get the optimal control sequence that could drive the systems to accomplish their tasks. %The model position at the end of the horizon is shown in Fig.\ref{figfinal}.
The training parameters and outcomes are summarized in (Tables \ref{d2c_comparison_table1}, \ref{d2c_comparison_table2}, \ref{params_comp}). We note that the training time is dominated by the Python simulation code, and can be made significantly faster with an optimized and parallellized code. Thus, we envisage this work as an initial proof of the concept for controlling material micro-structures. 

\begin{figure*}[!htpb]
\centering
\begin{multicols}{5}
   
      \subfloat[Initial State]{\includegraphics[width=1.2\linewidth]{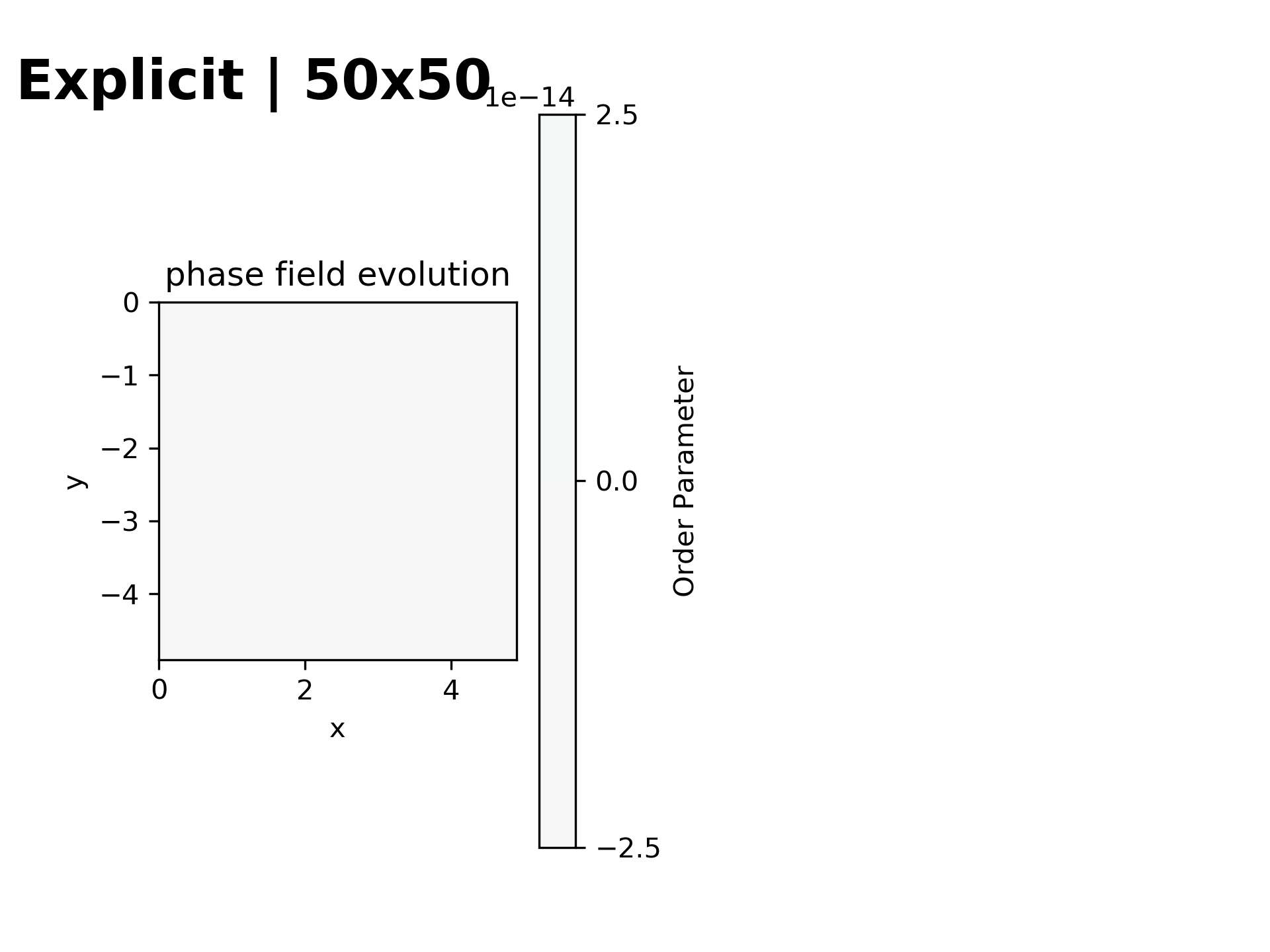}}    
       \subfloat[t=0.025s]{\includegraphics[width=1.2\linewidth]{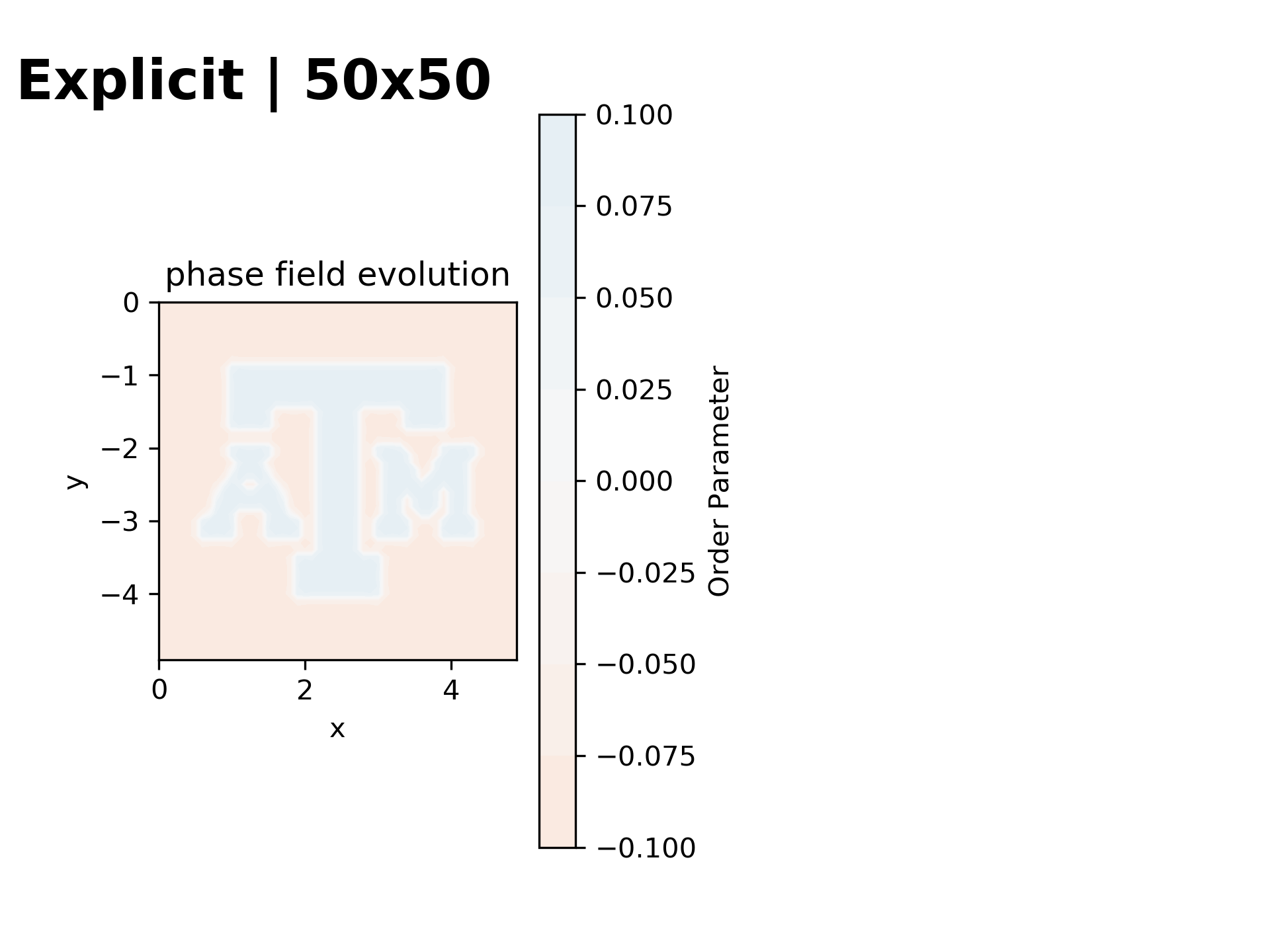}}
        \subfloat[t=0.075s]{\includegraphics[width=1.2\linewidth]{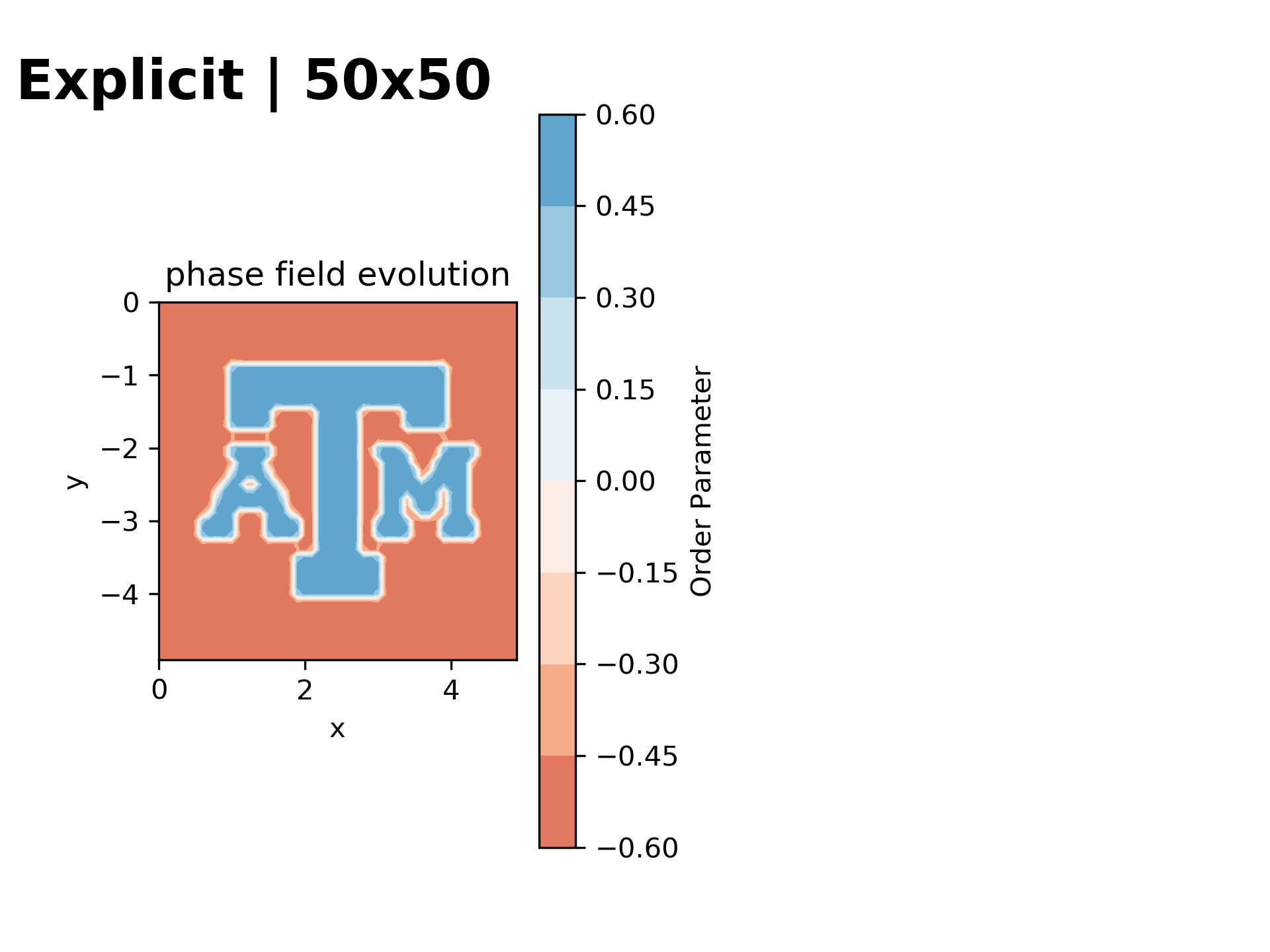}}
         \subfloat[t=0.125s]{\includegraphics[width=1.2\linewidth]{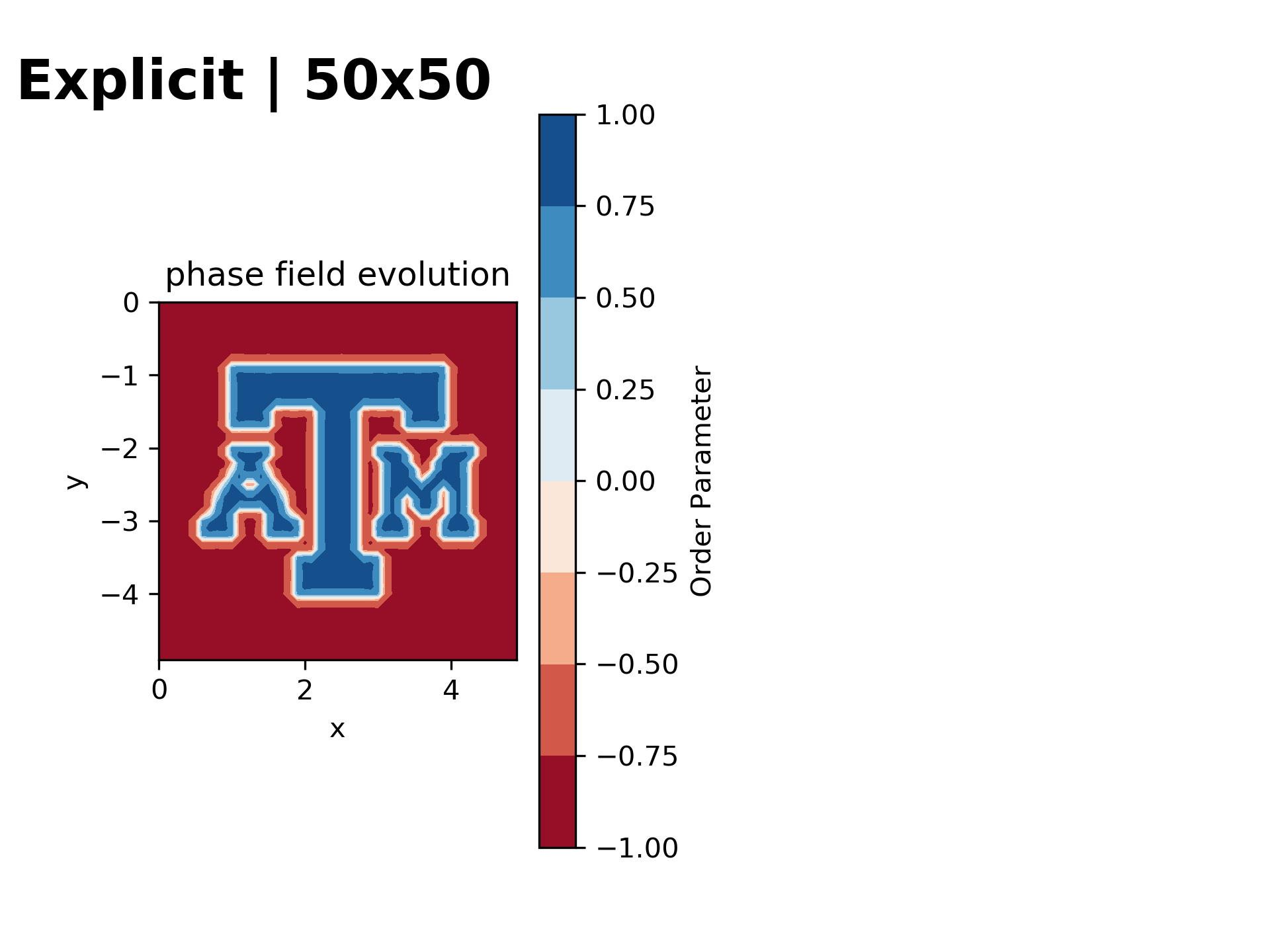}}
           \subfloat[Final State]{\includegraphics[width=1.2\linewidth]{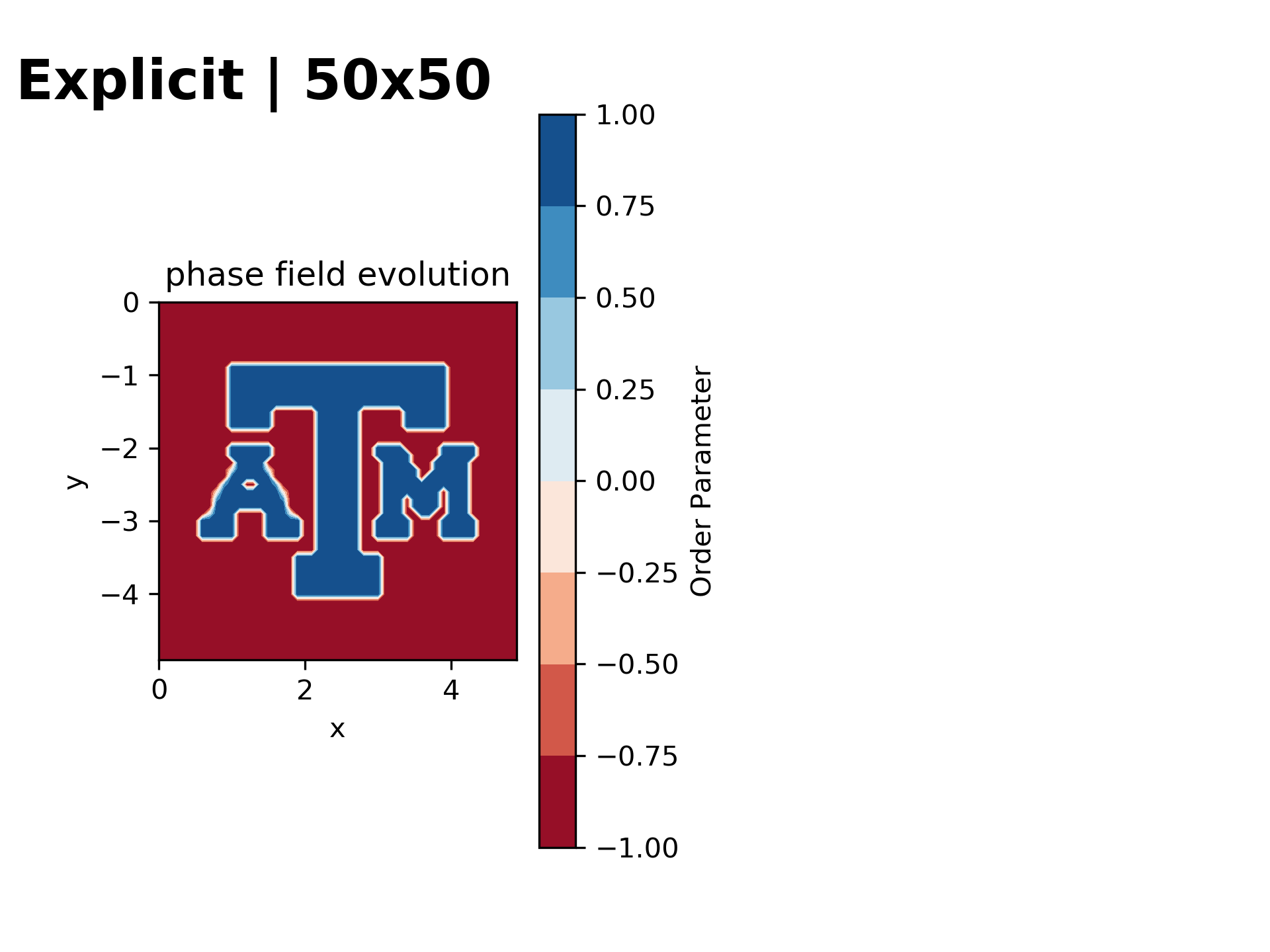}}\\

\end{multicols}

\caption{Open-loop optimal trajectory for goal-state III with the Allen-Cahn PDE as learned by the D2C algorithm.}
\label{d2c_2_atm}
\end{figure*}

The open-loop optimal trajectory learned by the D2C algorithm after convergence (for goal state-III, Allen-Cahn PDE) is shown in Figure \ref{d2c_2_atm}.
%\begin{figure}[!htb]
%     {\includegraphics[width=0.8\linewidth]{images/episodic_cost_OL_training.png}} 
%\caption{Episodic cost vs time taken during open-loop training}
%\label{cost}
%\end{figure}

% \begin{table}[h]
% \begin{threeparttable}
% \caption{D2C training parameters and outcomes}
% \label{train}
% %\begin{center}
% \setlength{\tabcolsep}{1.7mm}{
% \begin{tabular}{|c|c|c|c|c|c|}
% \hline
% System&Steps& Time-&Rollout& \multicolumn{2}{|c|}{Training time}\\
% &per&step&number&\multicolumn{2}{|c|}{(in sec.)} \\\cline{5-6}
% &episode&(in sec.)&& Open-& Closed-\\
% &&&&loop&loop\\
% \hline
% Material Model& 50 & 0.025&500&90601.33  &1908.7453\\
% \hline
% \end{tabular}
% }
% %w.r.t. 2-core CPU@2.9GHz, 12GB Memory
% %\end{center}
% \begin{tablenotes}
% \item[1] The open-loop training is run on a laptop with a 2-core CPU@2.9GHz and 12G RAM. No multi-threading at this point.
% \end{tablenotes}
% \end{threeparttable}
% \end{table}

\begin{table}[h]
%\begin{center}
\parbox{\linewidth}{
\caption{Comparison of the training outcomes of D2C with DDPG for the Allen-Cahn Model.}
\centering
\begin{tabular}{|c|c|c|c|}
\hline
& \multicolumn{2}{|c|}{\bf Training time (in sec.)}\\\cline{2-3}
{\bf Goal State}&{\bf D2C} &{\bf DDPG} \\
%{\bf State}& & \\
\hline
{\bf Goal-I(10X10)}  &15.83 &2567.71* \\
\hline
%{\bf Goal-II(20X20)}&12327.36& 55.206 &4419.36*\\
{\bf Goal-II(20X20)}& 55.206 &4419.36*\\
\hline
{\bf Goal-III(50X50)}&3289.354 &**\\
\hline

% {\bf 5X5} & 486.34 &143.78 &9647.83\\
% \hline
% {\bf 20X20}&18558.18& 535.84 &49061.74\\
% \hline
% {\bf 20X20}&18558.18& 535.84 &49061.74\\
% \hline
\end{tabular}
\label{d2c_comparison_table1}
\begin{tablenotes}
\item[1] The open-loop training is run on a laptop with a 2-core CPU@2.9GHz and 12GB RAM. No multi-threading at this point.
\item[2] * implies algorithm cannot converge to the goal state.
\item[3] ** implies system running out of memory at run-time.
\end{tablenotes}
%\end{center}
}
\end{table}
% \footnote{* implies algorithm cannot converge to the goal state}

\begin{table}[h]
%\begin{center}
\parbox{\linewidth}{
\caption{Comparison of the training outcomes of D2C with DDPG for the Cahn-Hilliard Model.}
\centering
\begin{tabular}{|c|c|c|}
\hline
% & \multicolumn{2}{|c|}{\bf Training time (in sec.)}\\\cline{2-3}
% {\bf Goal State}&{\bf D2C} &{\bf DDPG} \\
% %{\bf State}& & \\
% \hline
% {\bf Goal-I(10X10)}  &141.573 &3598.298* \\
% \hline
% %{\bf Goal-II(20X20)}&12327.36& 55.206 &4419.36*\\
% {\bf Goal-II(20X20)}& 5083.223 &9956.454*\\%7260.435*\\
% \hline
& \multicolumn{2}{|c|}{\bf Training time (in sec.)}\\\cline{2-3}
{\bf Goal State}&{\bf D2C} &{\bf DDPG}\\
\hline
{\bf Goal-I(10X10)} &141.573 &3204.121*\\
\hline
{\bf Goal-II(20X20)} &5083.223 &9956.454*\\
\hline

% {\bf 5X5} & 486.34 &143.78 &9647.83\\
% \hline
% {\bf 20X20}&18558.18& 535.84 &49061.74\\
% \hline
% {\bf 20X20}&18558.18& 535.84 &49061.74\\
% \hline
\end{tabular}
\label{d2c_comparison_table2}
\begin{tablenotes}
\item[1] The open-loop training is run on a laptop with a 2-core CPU@2.9GHz and 12GB RAM. No multi-threading at this point.
\item[2] * implies algorithm cannot converge to the goal state.
\end{tablenotes}
%\end{center}
}
\end{table}

\begin{table}[ht]
\caption{Parameter size comparison between D2C and DDPG}
\label{parasize}
\centering
\vspace{0.1in}
\begin{threeparttable}
\setlength{\tabcolsep}{0.6mm}{
\begin{tabular}{|c|c|c|c|c|}
\hline
System& No. of  & No. of & No. of  &No. of \\
&steps&actuators&parameters&parameters\\
&&&optimized&optimized\\
&&& in D2C& in DDPG\\
\hline
Goal-I&10  &200&2000&461901\\
\hline
Goal-II& 10 &800& 8000&1122501\\
\hline
Goal-III & 10&5000 & 50000&5746701\\
\hline
\end{tabular}
}
\vspace{0.1in}
\end{threeparttable}
\label{params_comp}
\end{table}

{\bf Testing Criterion:} For the closed-loop design, we proceed with the system identification and feedback gain design step of the D2C algorithm mentioned in the previous section to get the closed-loop control policy.  For testing, we compare the performance between the open-loop D2C control policy and the closed-loop D2C control policy under different noise levels. 
% We also compare the D2C closed-loop performance to the model based control. 
The open-loop control policy is to apply the optimal control sequence solved in the open-loop training step without any feedback. Hence, the perturbation drives the model off the nominal trajectory and increases the episodic cost as the noise level increases. Zero-mean Gaussian i.i.d. noise is added to every control channel at each step. The standard deviation of the noise is proportional to the maximum control signal in the designed optimal control sequence for D2C. 
% As for MBC, we use the optimal control values for the decoupled dynamics. 
% {\bf Model Predictive Control:}
% \textbf{Model Predictive Control(Reducing Horizon replanning):}

\begin{algorithm}
    \caption{Recursive MPC Algorithm}
    \label{alg3}
   {\bf 1)}  \textit{Given:} Initial state $\Phi_0$, time horizon $T$, cost $c(\Phi,U)=l(\Phi)+\frac{1}{2}rU^2$, and terminal cost $c_T(\Phi)$.\\
  {\bf 2)} Set $H=T$, $\Phi_i = \Phi_0$.\\
%   {\bf 3)} Solve the Riccati equations for each step along the nominal trajectory for feedback gain $\{ K_t \}_{t = 0}^{T-1}$.\\
%   {\bf 4)} Apply the closed-loop control policy,
  
\While{$H >0$}{
  \begin{enumerate}
        \item Solve the open-loop(deterministic) optimal \\control problem for initial state $\Phi_i$ and horizon \\$H$. Let the optimal sequence be $U^*=\{U_0, U_1,...U_{H-1}\}$.
        \item Apply the first control $U_0$ to the stochastic \\system,
        and observe the next state $\Phi_n$.
        \item Set $H=H-1, \Phi_i = \Phi_n$
\end{enumerate}
}
\end{algorithm}

\begin{figure}[!htb]

%\begin{multicols}{2}
\centering
\includegraphics[width=0.9\linewidth]{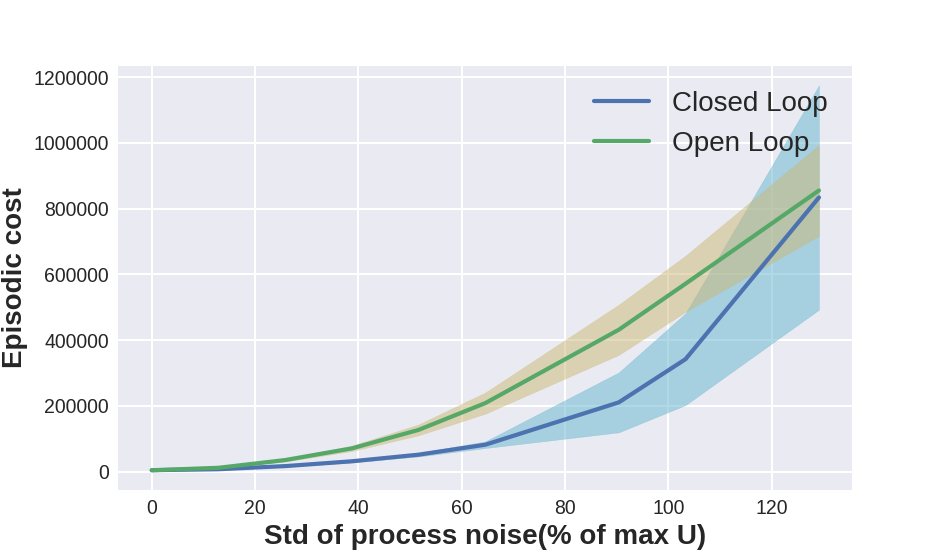}
%\subfloat[Initial state noise]{\includegraphics[width=\linewidth]{images/Comparison_delx1.png}}
%\end{multicols}

\caption{Performance comparison between D2C open-loop and closed-loop control policy}
\label{fig:clop}
\end{figure}

The feedback policy obtained appears to be near-optimal and highly robust for low noise levels (Fig \ref{fig:clop}), which is as expected from the $O(\epsilon^4)$-optimality of the algorithm used \cite{aistatsd2c2}. But when operating in high noise regimes, the robustness of the feedback policy quickly degenerates. 
Thus, to reduce the variance of the policy in these domains, we advocate the use of a rapid open loop solver along with replanning at every step, i.e. a \textbf{Model-Predictive Control} approach, to solve the RL problem, since MPC should be able to recover the global optimality for these problems \cite{mohamed2020near}.

\begin{figure}
    %\centering
    \includegraphics[width=0.9\linewidth]{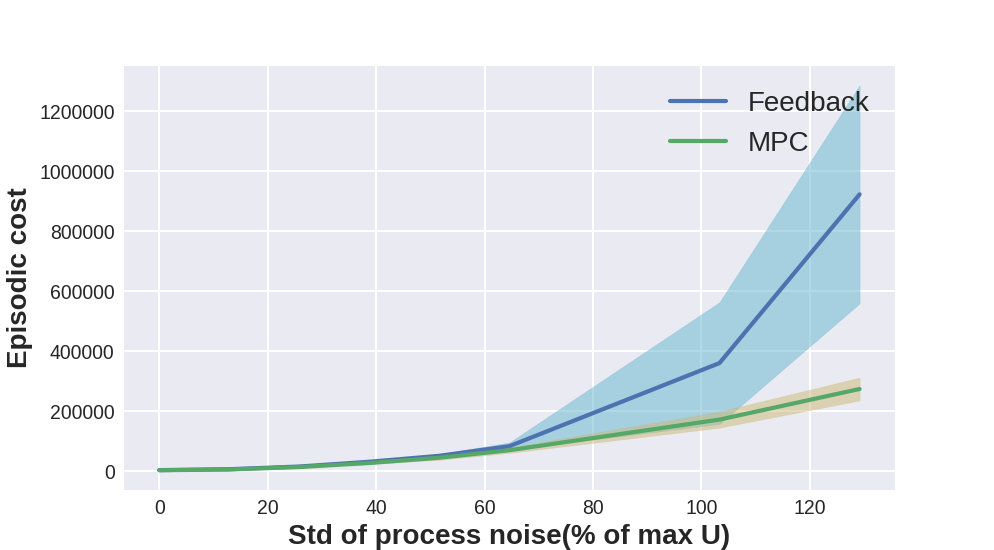}
    \caption{Comparing robustness to process noise between feedback-policy and recursive MPC.}
    \label{fig:replan_error}
\end{figure}

The recursive MPC algorithm implemented, which uses a fast and reliable local-planner, is summarised in Algorithm \ref{alg3}, and Figure \ref{fig:replan_error} compares the robustness to noise between the stochastic policy obtained from the recursive MPC with the decoupled closed-loop feedback policy at varying noise regimes.

It must be noted that the range of noise levels (at least up until about 100 \% of the maximum nominal control signal) that we are considering here is far beyond what is typically encountered in practical scenarios. As for the criterion for performance, we use episodic cost at each noise level. 100 rollouts are simulated at every noise level tested.

{\bf Discussion:} The total time comparison in Tables \ref{d2c_comparison_table1} and \ref{d2c_comparison_table2} shows that D2C learns the optimal policy significantly faster than DDPG. The primary reason for this disparity is the feedback parametrization of the two methods: the DDPG deep neural nets are complex parametrizations that are difficult to search over, when compared to the highly compact open-loop + linear feedback parametrization of D2C, as is apparent from Table \ref{parasize}. We observe from experiments that the DDPG algorithm is not able to converge to the goal state for the material system of state-dimension over ~5X5 discretization(25 states, 50 controls), while the D2C algorithm can quickly generate the closed-loop feedback solution even for the more complex systems. %We expect this difference to be even more apparent upon implementing multi-processing for the Python D2C code, which may lead to online learning for practical-scale microstructures. 

\section{CONCLUSIONS}

In this article, we have provided an overview of the modeling of multi-phase micro-structures in accordance with the Allen-Cahn and Cahn-Hiliard equations. We have also compared learning/ data-based approaches to the control of such structures, outlining their relative merits and demerits. In future work, we shall further develop a hybrid model-data based method which can exploit a priori knowledge of the dynamics to significantly reduce the computation time. We seek to implement a smarter actuator-selection scheme to reduce action space, which may improve viability of DDPG for higher order systems, while also significantly improving performance of D2C. We also plan to extend the methodology for higher-dimensional materials, leading up to an implementation on a realistic full-scale model.
\section{Acknowledgements}
This work was supported by the NSF CDS\&E program under grant number 1802867.
\bibliographystyle{IEEEtran}
\bibliography{IEEEabrv,ICRA_refs,TAC_refs}%

% \begin{thebibliography}{}
% \setlength{\itemindent}{-\leftmargin}
% %\makeatletter\renewcommand{\@biblabel}{}\makeatother
% % approximate DP
% @book{bensoussan,
%   title={Representation and control of infinite dimensional systems},
%   author={Bensoussan, Alain and Da Prato, Giuseppe and Delfour, Michel C and Mitter, Sanjoy K},
%   volume={1},
%   year={1992},
%   publisher={Birkh{\"a}user Boston}
% }

% \end{thebibliography}

\newpage
% \input{Appendix.tex}

%D2C : NEAR OPT, OL-Grad Desc and ILQR, CL {Done}

%ADD stability of the CS plots {Done}

% SPLIT OPEN-LOOP THEORY TO GD-based AND ILQR-based? {Done}
%Global optimal proof? {Not needed}

%Sys ID by LLS-CD expansion needed?? {Done}

%O(e) optimality proofs needed?? {Not needed}

%REVIEW: In total, the Algorithm 1 appears as an iterative procedure of state discretization, obtaining set of local LTV models, and designing standard LQR regulator by Riccati approach. What is here special or new as for control design??  {We are not calling it unique, just comparing methods and suggesting such an approach to be more practical}

%JOURNAL VERSION
% -Re add gradient descent comparisons and plots
% -Add details on workings of AC, CH equation(effect of transport)
% -Add a section on replanning at every step, maybe do a plot on replanning times per iteration
% -Add a section on ILQR (Tassa IROS 2012?)

\end{document}